\documentclass[twocolumn,prl,tightenlines,superscriptaddress,showpacs]{revtex4-2}

\usepackage{amsmath}
\usepackage{amssymb,amsfonts,latexsym}
\usepackage{bm}
\usepackage[mathcal]{euscript}
\usepackage{graphicx}
\usepackage{epsfig}

\usepackage[svgnames]{xcolor}
\usepackage{xfrac}
\usepackage{mathrsfs}

\usepackage[colorlinks,linkcolor=MediumBlue,citecolor=MediumBlue,urlcolor = MediumBlue]{hyperref}
\usepackage{url}

\allowdisplaybreaks

\newcommand{\colrev}[1]{{\color{black} #1}}

\newcommand{\rmd}{{\rm d}}
\newcommand{\calP}{{\cal P}}
\newcommand{\n}{\hat{\bm n}}
\newcommand{\mv}{\langle v \rangle}
\newcommand{\mvsq}{\langle v^2 \rangle}
\newcommand{\br}{\bar{\rho}}

\begin{document}

\title{Emergent Organization and Polarization due to Active Fluctuations}

\author{Beno\^{\i}t Mahault}\email{benoit.mahault@ds.mpg.de}
\affiliation{Max Planck Institute for Dynamics and Self-Organization (MPIDS), 37077 G\"ottingen, Germany}

\author{Prakhar Godara}
\affiliation{Max Planck Institute for Dynamics and Self-Organization (MPIDS), 37077 G\"ottingen, Germany}

\author{Ramin Golestanian}\email{ramin.golestanian@ds.mpg.de}
\affiliation{Max Planck Institute for Dynamics and Self-Organization (MPIDS), 37077 G\"ottingen, Germany}
\affiliation{Rudolf Peierls Centre for Theoretical Physics, University of Oxford, Oxford OX1 3PU, United Kingdom}

\date{\today}

\begin{abstract}
We introduce and study a model of active Brownian motion with multiplicative noise describing fluctuations in the self-propulsion or activity. We find that the standard picture of density accumulation in slow regions is qualitatively modified by active fluctuations, as stationary density profiles are generally not determined only by the mean self-propulsion speed landscape. As a result, activity gradients generically correlate the particle self-propulsion speed and orientation, leading to emergent polarization at interfaces pointing either towards dense or dilute regions depending on the amount of noise in the system. 
We discuss how active noise affects the emergence of motility induced phase separation. Our work provides a foundation for systematic studies of active matter self-organization in the presence of activity landscapes and active fluctuations.
\end{abstract}

\maketitle


Active matter refers to a broad class of systems driven out of thermal equilibrium at the level of their microscopic constituents.
Although detailed balance can be broken locally in various ways, including generation of local stresses~\cite{Prost2015NatPhys,Doostmohammadi2018NatComm}, 
local production and consumption of chemicals~\cite{Golestanian2019phoretic},
non-reciprocity in the interactions~\cite{Soto:2014,Saha2019,Saha2020PRX,You2020PNAS,DadhichiPRE2020,Fruchart2021Nat},
or growth~\cite{DellArciprete2018NatComm,You2018PRX,Hartmann2019NatPhys}, systems composed of self-propelled particles play a central role
in active matter studies~\cite{ABPreview2012,MarchettiRMP2013,Cates2015MIPS,Gompper2020IOP,ChateDADAM2020}.
An interesting feature of the latter case is that activity trivially couples to the particles motion, 
so that it can be used as a way to control the spatial dynamics of active systems with promising practical applications~\cite{Gao2014ACS,Ghosh2020Nano}.
Spatial segregation of active particles can be achieved at the individual particle level, e.g.\ by imposing spatially varying self-propulsion speed~\cite{Lozano2016NatComm,Arlt2018NatComm,frangipane2018Elife,Soker2021PRL},
\colrev{or at the collective level by engineering quorum-sensing-type interactions triggering motility inhibition
in single~\cite{Bauerle2018NatComm,Lavergne2019}, and multi-component~\cite{Curatolo2020Natphys} suspensions.}

The physical principles leading to such controls apply to a wide class of systems and can thus be understood from minimal models, 
which most often deal with Active Brownian Particles (ABPs)~\cite{ABPreview2012}.
ABPs are generally studied in the overdamped regime under the assumption that their self-propulsion speed relaxes on infinitesimal timescales,
so that it can be considered constant, and the stochastic dynamics of the self-propulsion velocity reduces to rotational diffusion.
Actual active particles, on the other hand, self-propel as a result of intricate processes involving nontrivial timescales~\cite{Elgeti2015review,RG:2009}, 
and usually evolve inside complex and noisy environments~\cite{BechingerRMP2016}.
Moreover, individual measurements of active self propulsion velocity often shows it to be a dynamical fluctuating quantity~\cite{BERG1972Nature,Magariyama1995BiophysJ,Paxton2004ACS,Dreyfus2005Nature,Howse2007PRL,Corkidi2008Biochem,Thutupalli2011NJP,Grosjean2016PRE,Turner2016BioJ,Fragkopoulos2021}, which is not a surprise, given what we expect from mechanistic descriptions of self-propulsion.
\colrev{ 
Phoretic colloids, for instance, are typically driven at their surface by a chemical reaction whose product density is a fluctuating quantity~\cite{RG:2005,RG:2009}.
Low Reynolds swimmers, in turn, must self-propel by performing nonreciprocal cycles that may be described as stochastic transitions between internal states~\cite{Najafi:2004}.
In both cases, fluctuations of the motility mechanism naturally lead to fluctuations in the self-propulsion speed.
When the source of activity is inhomogeneous, e.g. in the presence of chemical product density or swimming medium viscosity gradients, such fluctuations can moreover vary in space.
}

\colrev{Fluctuations of the active velocity, which hereafter we refer to as {\it active fluctuations}, must by symmetry carry independent contributions 
along the directions parallel and perpendicular to the self-propulsion direction.}
Unlike thermal noise, active noise is therefore multiplicative by design.
Despite a few studies characterising statistical properties of active fluctuations~\cite{Schienbein1993BMB,peruaniPRL2007,Romanczuk2011PRL,ABPreview2012,Chaudhuri2014PRE,Caprini2022JCP}, 
little is known about how the latter can in turn affect the spatial organization of the particles.

\colrev{Here, we propose a formulation of fluctuating active Brownian motion 
for which the particle self-propulsion is selected by a generalized velocity potential and fluctuates both in norm and direction.
Considering the Fokker-Planck equation describing this fluctuating active dynamics, 
our analysis reveals that the expression of the nonequilibrium current in the hydrodynamic regime depends tightly on the details of the self-propulsion mechanism, while it satisfies the same symmetries as for constant speed ABPs.
Focusing on the case where the self-propulsion velocity derives from a quadratic potential,
we moreover characterize the steady-state properties of the model in the presence of spatially varying activity. We find that density profiles created by activity landscapes are not only set by the mean particle self-propulsion speed but also by its higher order moments.
We find that activity gradients generically correlate the particles self-propulsion speed and orientation, and lead to an emergent polarization whose direction can point up or down the gradient depending on the model parameters.}
We discuss the consequences of the above features on the emergence of motility induced phase separation~\cite{Cates2015MIPS}.
Our work demonstrates that active fluctuations can be used as a way to control the spatial organization of active matter,
and leads to conclusions in sharp contrast with existing results obtained in the limit of constant speed ABPs.


\colrev{
\paragraph{ABPs with active fluctuations.---}
Contrary to thermal noise, the symmetries of active noise are dictated by that of the particle's self-propulsion velocity $\bm v$.
Fluctuations of the norm $v \equiv |\bm v|$ and direction $\n \equiv \bm v / v$ of the self-propulsion are thus generally decoupled.
Therefore, we consider a model of overdamped ABPs in dimension $d \ge 2$ described by the following Langevin equations
\begin{subequations}
\label{eq_Langevin}
\begin{align}
\label{eq_Langevin_r}
\dot{\bm r} & = \bm v + \frac{1}{2}\nabla D_t(\bm r) + \sqrt{2 D_t(\bm r)}\bm \xi_t(t) ,\\
\label{eq_Langevin_v}
\dot{v} & = -\partial_v W(\bm r,v) + \sqrt{2D_v(\bm r)} \xi_v(t), \\
\label{eq_Langevin_n}
\dot{\n} & = \sqrt{2 D_r(\bm r)} {\bm P}^{\perp}(\hat{\bm n}) \cdot \bm \xi_r(t) ,
\end{align}
\end{subequations}
where all model parameters may depend on the particle position $\bm r$ while
the unit variance, uncorrelated, Gaussian white noises $\bm \xi_t$, $\xi_v$, and $\bm \xi_r$ are interpreted in the Stratonovich sense. 
Equation~\eqref{eq_Langevin_r} describes the spatial dynamics of the particle with translational diffusivity $D_t(\bm r)$.
It moreover comprises a term compensating for the noise-induced drift that arises due to the multiplicative nature of the translational noise.
The r.h.s.\ of Eq.~\eqref{eq_Langevin_v} gathers two contributions.
The first one corresponds to a deterministic active force that derives from an effective potential $W(\bm r,v)$,
while the second contribution results from active fluctuations which are parameterized by the coefficient $D_v(\bm r)$.
Finally, Eq.~\eqref{eq_Langevin_n} sets the orientational dynamics of self-propulsion with associated diffusivity $D_r(\bm r)$, where $P^{\perp}_{ij}(\n) \equiv \delta_{ij} - \hat{n}_i \hat{n}_j$ denotes the projector orthogonal to $\n$.
Note that Eqs.~(\ref{eq_Langevin_v},\ref{eq_Langevin_n}) can equivalently be expressed in terms of the velocity $\bm v$.
Keeping the Stratonovich interpretation of the noise, this leads to~\footnote{See Supplementary Material at [url].}
\begin{equation*}
\dot{\bm v} = -[\partial_v W +(d-1)\sqrt{D_v D_r}]\n + \sqrt{2}\bm \Sigma(\bm v) \cdot \bm \xi(t),
\end{equation*}
with $\bm \Sigma(\bm v) \equiv \sqrt{D_v}\n\n + v \sqrt{D_r}{\bm P}^{\perp}(\hat{\bm n})$.
Therefore, due to the decoupling between the speed and orientational degrees of freedom, the noise on the self-propulsion velocity is generally multiplicative.
}

\colrev{
Denoting ${\cal P}({\bm r},v,\n,t)$ as the single-particle distribution, it is straightforward to show that its dynamics follows
\begin{align}
\partial_t {\cal P} = -\nabla \cdot \bm J_{\bm r}  - v^{1-d}\partial_v\left(v^{d-1} J_v \right) + D_r\nabla_{\hat{\bm n}}^2 {\cal P},
\label{eq_Kramers}
\end{align}
where $\nabla_{\hat{\bm n}}^2 \equiv P_{ij}^\perp(\hat{\bm n})\partial^2_{\hat{n}_i \hat{n}_j} - (d-1)\hat{n}_i\partial_{\hat{n}_i}$ denotes the spherical Laplacian (summation over repeated indices is assumed), 
and the dependencies on $\bm r$, $v$ and $\n$ are now implicit.
The translational and speed currents are respectively given by $\bm J_{\bm r} \equiv (\bm v - D_t\nabla){\cal P} $ and $J_v \equiv -\partial_v W_{\rm eff}{\cal P}  - D_v \partial_v {\cal P} $,
where the effective potential is defined as $W_{\rm eff} \equiv W + (d-1)D_v \ln v$.
}

When the diffusivities and $W$ are independent of $\bm r$, 
steady state solutions of Eq.~\eqref{eq_Kramers} can be factorized and the speed distribution takes the Boltzmann-like form~\cite{Romanczuk2011PRL}
\begin{equation} \label{eq_PH}
{\cal P}_{\rm H}(v) = Z^{-1}e^{-W_{\rm eff}(v)/D_v} = Z^{-1}v^{1-d}e^{-W(v)/D_v} ,
\end{equation}
with $Z \equiv \int_0^\infty {\rm d}v \, e^{-W(v)/D_v}$ ensuring normalization.


\paragraph{Moment expansion.---}
If the coefficients of Eq.~\eqref{eq_Kramers} are spatially dependent, on the other hand, 
the active drift in $\bm J_{\bm r}$ coupling self-propulsion speed and orientation prevents any factorization of ${\cal P}$. 
However, the large-scale and long-time dynamics of the system is generally well captured by means of an expansion of Eq.~\eqref{eq_Kramers} in the moments of $\bm v$.
Given the symmetries of the problem, we shall not directly consider moments associated with $\bm v$,
but rather those related to the corresponding speed $v$ and orientation $\hat{\bm n}$.
For the following discussion, we thus define the average: $\rho \langle \cdot \rangle \equiv \int v^{d-1}{\rm d} {v} {\rm d}\n \, (\cdot) {\cal P}({\bm r},v,\n,t)$,
where $\rho({\bm r},t) = \rho\langle 1 \rangle$ is the local particle density normalized to unity.
Deriving the equations for the moments (details in Appendix A), we find that the nature and number of terms they involve depend closely on the specific form of $W$.
It is moreover clear that the unique slow mode of the dynamics is the conserved particle density $\rho$, which satisfies $\partial_t \rho = -\nabla\cdot \bm J$
with $\bm J = \rho \langle v \n \rangle - D_t \nabla\rho$.
Therefore, in the long-time and large-scale limit the dynamics of all higher order --in speed and orientations-- moments can be enslaved to $\rho$.

In order to keep the presentation simple while retaining relevant features of active fluctuations, 
we consider the quadratic form 
$ W({\bm r},v) = \tfrac{\mu(\bm r)}{2} \left[ v - v_0(\bm r)\right]^2 $,
where $\mu$ and $v_0$ are function of space and assumed positive.
\colrev{With this choice of the potential, the velocity dynamics has four parameters which can all be experimentally evaluated from individual particle tracking.
$\tau_r \equiv ((d-1)D_r)^{-1}$ sets the typical timescale of rotational diffusion, while $\mu^{-1}$ controls speed relaxation.
Assuming a homogeneous system, both can be measured from the steady-state autocorrelation of the particle self-propulsion direction and speed.
Namely, $\langle \n(t) \cdot \n(t + \tau)\rangle = e^{-\tau/\tau_r}$ and $\langle v(t) v(t + \tau)\rangle = \langle v^2 \rangle e^{-\mu \tau}$.
Moreover, $v_0$ and $D_v$ can be obtained from the full speed distribution~\eqref{eq_PH},
or simply from its first two moments
\begin{equation} \label{eq_speed_moments_quadratic}
\mv_{\rm H} = v_0 + \frac{D_v}{\mu}\gamma, \qquad  \mvsq_{\rm H} = \frac{D_v}{\mu} + v_0\mv_{\rm H},
\end{equation}
where the subscript H refers to averages computed with the distribution~\eqref{eq_PH}, 
while $\gamma = \sqrt{2\mu/(D_v\pi)}f(\mu v_0^2/D_v)$ and $f(x) \equiv e^{-x/2}/[1 + {\rm Erf}(\sqrt{x/2})]$.
We note that the limit of constant speed ABPs is recovered for $\mu \to \infty$, 
such that $\mv_{\rm H}$ instantly relaxes to $v_0$.
On the other hand, taking $D_v$ and $\mu$ finite with $v_0 = 0$ amounts to a variant of Active Ornstein-Uhlenbeck particles~\cite{Martin2021PRE} 
where speed and orientation fluctuations are decoupled.
}

We show in Appendix A that up to ${\cal O}(\nabla^2)$ terms the self-propulsion speed moments are given by $\langle v^k \rangle = \langle v^k \rangle_{\rm H}$ 
where $\calP_{\rm H}(\bm r,v)$ now varies in space as a result of the spatial dependencies of $W$ and $D_v$.
Similarly, we find that at this order in gradients the polarity moments can be expressed in terms of $\nabla (\rho \langle v^k \rangle)$ with $k \ge 1$.
Replacing the relevant expressions into the particle density current, we finally obtain in the hydrodynamic limit:
\begin{equation} \label{eq_J_quadratic}
\bm J = - D_t \nabla \rho - \frac{\tau_r}{d(1+\alpha)}\left[ \nabla \left(\rho \langle v^2 \rangle\right) + \alpha v_0 \nabla \left(\rho \langle v \rangle \right)\right] ,
\end{equation}
where $\alpha \equiv \mu \tau_r$.
For a quadratic potential $W$, the density dynamics is therefore uniquely determined by the first two self-propulsion speed moments. 
For $v_0 > 0$ and a fast active speed dynamics ($\alpha \gg 1$), 
the dependency of $\bm J$ in $\langle v^2 \rangle$ moreover drops out so that its expression reduces to that of constant speed ABPs~\cite{SchnitzerPRE1993}.
Below, we discuss the consequences of finite $\alpha$ values on the steady-state density.
We moreover present parameter free comparisons of our results with simulations of the Langevin system~\eqref{eq_Langevin} in three dimensions; see~\cite{Note1} for more details about numerical methods.


\paragraph{Inhomogeneous steady-states.---}
For flux-free boundary conditions, the stationary density profile $\rho_s(\bm r)$ is simply obtained from the condition $\bm J = \bm 0$, leading to
\begin{equation} \label{eq_ss_rho_quadratic}
\nabla \ln\rho_s = -\frac{\nabla \langle v^2 \rangle  + \alpha v_0 \nabla \langle v \rangle}{ v_D^2(1 + \alpha) + \langle v^2 \rangle + \alpha v_0 \langle v \rangle} ,
\end{equation}
where we have defined $v_D \equiv (d D_t\tau_r^{-1})^{1/2}$ as the velocity scale built from the translational and rotational diffusivities.
Equation~\eqref{eq_ss_rho_quadratic} predicts that spatially varying activity resulting in non-vanishing 
gradients of the self-propulsion speed moments leads to inhomogeneous density profiles.
From the expressions of the moments in Eq.~\eqref{eq_speed_moments_quadratic}, particle segregation can thus be achieved 
independently by imposing gradients of $v_0$, $D_v$ or $\mu$. 
Conversely, the translational diffusion coefficient only appears in Eq.~\eqref{eq_ss_rho_quadratic} through $v_D$, 
so that its spatial variations will not qualitatively affect $\rho_s$.
For simplicity, we thus neglect its contribution for the remaining of this discussion.

\begin{figure}[t!]
	\includegraphics[width = 0.99\columnwidth]{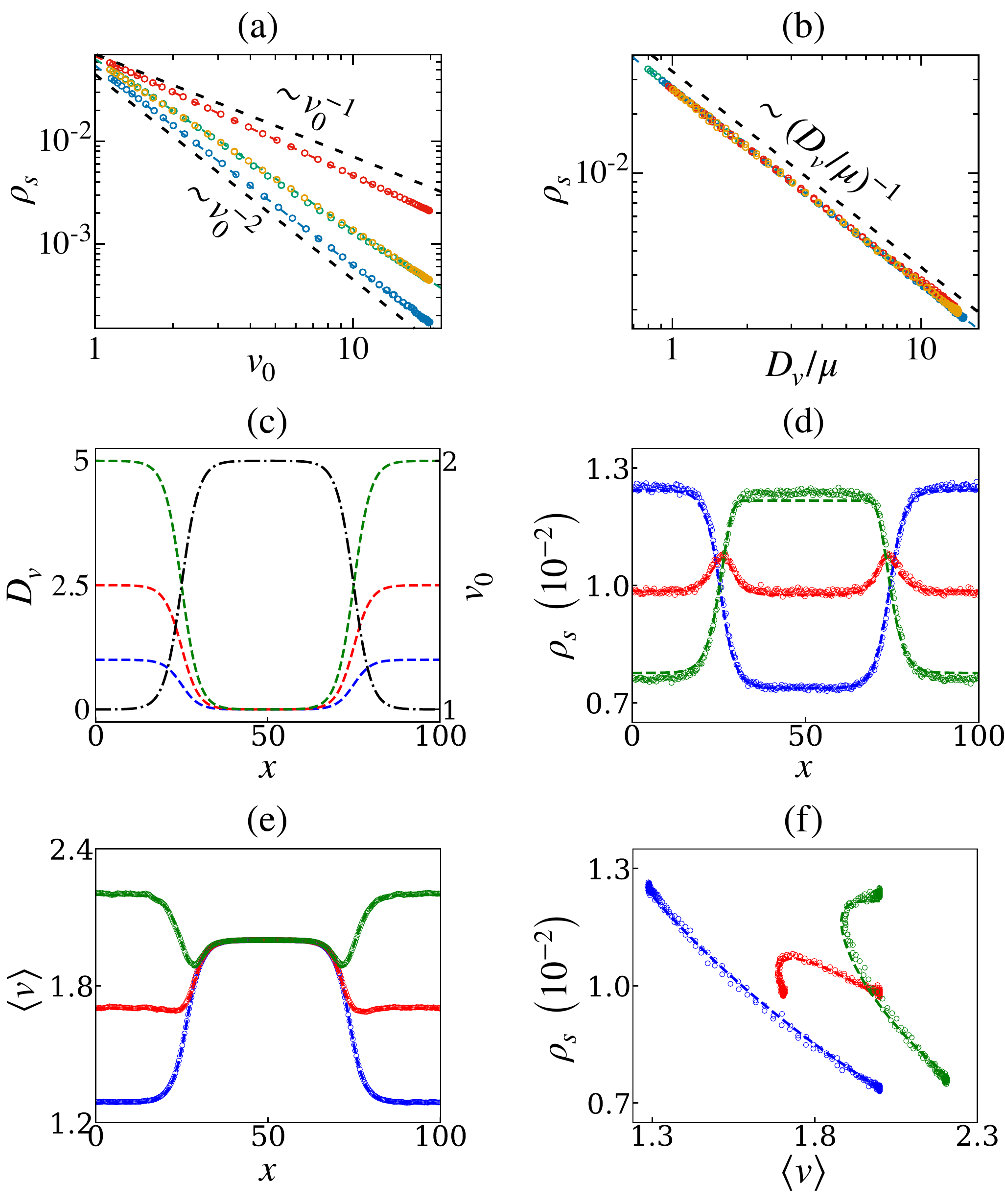}
	\caption{Steady state density with spatially varying activity.
	(a) Mean stationary density as function of $v_0$ for $\mu v_0^2 \gg D_v$
	; the colors respectively correspond to $(\mu,D_r) = (1,1)$ (green), (1,10)(blue), (10,1)(red) and (10,10)(yellow).
	(b) Mean stationary density as function of $D_v/\mu$ for $\mu v_0^2 \ll D_v$, the color code is same as (a).
	In (a,b) the data have been shifted vertically for clarity. 
	(c) Imposed one dimensional $v_0$ (black dash-dotted line) and $D_v$ (colored dashed lines) 
	profiles splitting the space in two nearly homogeneous regions with different activities. 
	(d,e) The mean density(d) and particle speed(e) profiles corresponding to (c).
	(f) The relation $\rho(\langle v \rangle)$ corresponding to (d,e).
	In (d-f) $\mu = D_r = 1$, the $v_0(x)$ profile is unchanged and the color codes the maximal value of $D_v$ in each case.
	In all panels except (c) open circles show results obtained from Langevin simulations while dashed lines indicate the corresponding theoretical predictions.
	}
	\label{fig1}
\end{figure}

Setting $v_D = 0$, the rotational diffusivity $D_r$ appears in the expression of $\rho_s$ only through the parameter $\alpha$.
Spatial variations of $D_r$ alone are therefore unable to generate density gradients, 
in agreement with usual considerations~\cite{Cates2015MIPS} and recent experimental results~\cite{Fernandez2020NatCom}.
However, the value of $\alpha$ sets the scaling of the steady state density with the speed moments.
For fast speed dynamics $(\alpha \gg 1)$ Eq.~\eqref{eq_ss_rho_quadratic} leads to the standard relation 
$\rho_s \sim \langle v \rangle^{-1}$~\cite{SchnitzerPRE1993},
while in the opposite case of fast rotational diffusion $(\alpha \ll 1)$ we instead get $\rho_s \sim \langle v^2 \rangle^{-1}$.
Assuming $\alpha$ to be uniform, we moreover obtain from Eq.~\eqref{eq_ss_rho_quadratic} in the active force, or noise dominated, regimes
\begin{subequations} \label{eq_simplifies_rho_sols}
\begin{align} \label{eq_simplifies_rho_sols_v0}
\rho_s &\sim v_0^{-(2+\alpha)/(1 + \alpha)} & &(D_v \ll \mu v_0^2) ,\\
\rho_s &\sim \langle v \rangle^{-2} \sim \left( D_v/\mu \right)^{-1} & &(D_v \gg \mu v_0^2) .
\end{align}
\end{subequations}
For low noises, the steady-state density therefore scales algebraically with the mean particle speed with an exponent set by
the ratio between rotational diffusion and speed relaxation timescales as shown in Fig.~\ref{fig1}(a).
On the contrary, when active noise dominates Fig.~\ref{fig1}(b) shows that the effect of rotational noise disappears as $\rho_s$ always scales as the inverse of $D_v/\mu$. 

From Eq.~\eqref{eq_ss_rho_quadratic}, for $D_v$ and $\mu v_0^2$ of similar order the stationary density is no more enslaved to the mean particle speed, as 
gradients of $\langle v^2 \rangle$ can compete with that of $\langle v \rangle$.
This feature allows for counter intuitive behavior, as we show by considering the following illustrative setup: 
we partition the space into two distinct sub-regions 1 and 2 in which active particles experience different uniform values of 
$v_0 = v_{1,2}$ and $D_v = D_{v_{1,2}}$ (see Fig.~\ref{fig1}(c)).
We denote $\tilde v = v_1 - v_2$ and $\tilde D_v = D_{v_{1}} - D_{v_{2}}$, and consider the case $\tilde v  \tilde D_v < 0$.
$\tilde v$ is kept fixed, and we vary $\tilde D_v$.
For sufficiently small $|\tilde{D}_v|$, $\rho_s$ is maximal in the region with the smallest $v_0$, 
as generally the case for constant speed ABPs~\cite{SchnitzerPRE1993,Cates2015MIPS} (see blue lines in Figs.~\ref{fig1}(c,d)).
Increasing $|\tilde{D}_v|$ progressively leads to an inversion of the density profile, such that for large $|\tilde{D}_v|$ 
particles instead accumulate on average in the region where $v_0$ is largest (green lines in Figs.~\ref{fig1}(c,d)).
The density inversion partly follows the behavior of the mean speed $\langle v \rangle$, 
also affected by $|\tilde{D}_v|$ (see Eq.~\eqref{eq_speed_moments_quadratic} and Fig.~\ref{fig1}(e)).
While $\rho_s$ globally remains largest in small $\langle v \rangle$ regions,
a direct inspection of the $\rho_s(\langle v \rangle)$ curves reveals that when $\nabla v_0$ and $\nabla D_v$ are both nonzero, 
$\rho_s$ can locally {\it increase} with $\langle v \rangle$ (Fig.~\ref{fig1}(f)), which is something impossible in the absence of active fluctuations.

\begin{figure}[t!]
	\includegraphics[width = 0.99\columnwidth]{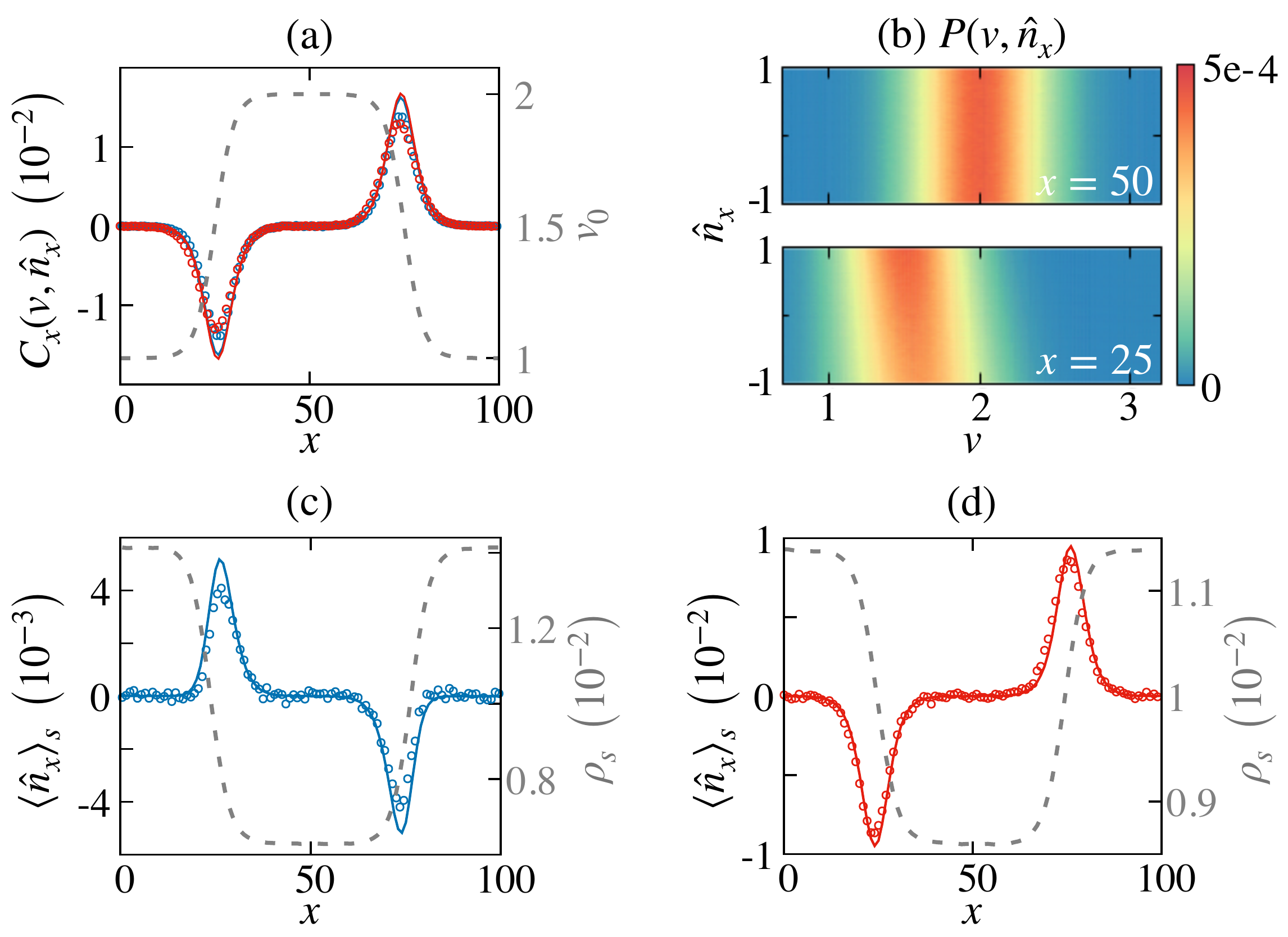}
	\caption{Speed-order correlations and local polarization.
	(a) The steady state correlation function $C_x(v,\hat{n}_x)$ in presence of spatially varying activity $v_0(x)$ (grey dashed lines) 
	as well as uniform $D_v = 0.1$ and $\mu = D_r = 1$ with $D_t = 0.1$(blue) and 1(red).
	(b) The joint speed-orientation distribution associated with (a) for $D_t = 0.1$, 
	at the center of the domain ($x = 50$) and at the interface between the two regions with different activities ($x=25$).
	(c,d) The local averaged polarization profiles corresponding to (a) with the same color code, while the dashed grey line shows the corresponding stationary density profiles.
	In (a,c,d) open circles show the simulations data, while continuous lines indicate theoretical predictions.
	}
	\label{fig2}
\end{figure}

\paragraph{Orientation-speed correlations and emergent polarization.---}
The phase space distribution solving Eq.~\eqref{eq_Kramers} with spatially varying coefficients cannot be factorized
due to the coupling between particle motion and activity (Fig.~\ref{fig2}(b)).
We now discuss the consequences of such feature on the particles orientational dynamics in the presence of activity landscapes.
In order to avoid dealing with lengthy expressions, we focus on the limiting case $\mu v_0^2 \gg D_v$,
but the results presented below hold in a more general context.
General expressions are presented in~\cite{Note1}.

Correlations between the speed and the orientation of the active particles are quantified 
considering the connected moment $\bm C(v,\hat{\bm n}) \equiv \langle v \hat{\bm n}\rangle - \langle v \rangle\langle \hat{\bm n}\rangle$,
fo which we find
\begin{align} \label{eq_Correlations}
\bm C(v,\hat{\bm n}) & = -\frac{\tau_r v_0 \nabla v_0}{d(1+\alpha)} \qquad (\mu v_0^2 \gg D_v), 
\end{align}
showing that $\bm C$ is nonzero in regions of non-vanishing activity gradients, as ilustrated in Fig.~\ref{fig2}(a).
Naturally, speed-orientation correlations are moreover suppressed in the cases of fast rotational and speed relaxation, respectively for $\tau_r \to 0$ and $\alpha \gg 1$.

Spatially varying activity moreover spontaneously generates local orientational order.
Namely, using the zero flux solution~\eqref{eq_ss_rho_quadratic} and the expression of the polarization given in Appendix A, we find in stationary state 
\begin{align} \label{eq_sol_n_ss}
\langle \hat{\bm n} \rangle_s & = \frac{\tau_r}{d}\frac{v_0^2 - v_D^2(1+\alpha)}
{(1+\alpha)(v_D^2 + v_0^2)}\nabla v_0 \quad (\mu v_0^2 \gg D_v), 
\end{align}
Equation~\eqref{eq_sol_n_ss} highlights two mechanisms at the origin of polar order in regions of nonzero activity gradients.
For fast speed relaxation ($\alpha \gg 1$), the steady-state active flux $\langle v \hat{\bm n} \rangle_s \sim v_0 \langle \hat{\bm n} \rangle_s$ 
must compensate the diffusive flux $-D_t\nabla\rho_s$
so that $\langle \hat{\bm n} \rangle_s$ points towards slow (or dense) regions (see Fig.~\ref{fig2}(d))~\cite{Fischer2020PRE,Row2020PRE,Soker2021PRL}.
This gradient alignment mechanism is consistent with the polarization generally observed at interfaces of repelling ABPs in the phase separated regime~\cite{Lee2017SoftMat,Solon2018NJP,Omar2020PRE},
where in this case the mean-field spatial diffusion results from particle collisions~\cite{Bialke2013EPL}. 
On the other hand, for self-propulsion speeds $v_0$ such that $v_0^2 > v_D^2(1+\alpha)$ 
Eq.~\eqref{eq_sol_n_ss} predicts a reversal of the mean polarization direction 
towards fast (or dilute) regions (see Fig.~\ref{fig2}(c)).
Indeed, neglecting the effect of positional diffusion particles crossing an interface separating two regions with different activities will be on average slower
--and thus stay longer at the interface-- if they come from the slower (or denser) region.  
Consequently, the local polarization will in this case be oriented towards the most dilute region.

\colrev{
\paragraph{Motility Induced Phase Separation (MIPS).---}  
The origin of MIPS can be understood at the mean field level from a mapping to a dynamics with quorum sensing interactions, 
where the particles' motility effectively depends on their local density~\cite{Cates2015MIPS,Bialke2013EPL}.
In the case where active particles slow down sufficiently fast as they reach denser regions, homogeneous systems may indeed undergo an instability which marks the onset of phase separation.
As the above derivation considers general spatial dependencies of the model parameters, it allows us to study how active fluctuations can trigger or influence the onset of MIPS.
Here, we assume $\rho$ to be sufficiently large such that its fluctuations can be neglected~\cite{Solon2015EPJE}. 
Consequently, we consider Eq.~\eqref{eq_J_quadratic} with all coefficients being functions of $\varrho \equiv \int\rmd^d\bm r'\, K(|\bm r - \bm r'|)\rho(\bm r')$, where $K$ is a (normalized) short-ranged interaction kernel.
Expanding $\varrho$ in the gradients of $\rho$, the active noise dynamics with quorum sensing interactions
maps to Active Model B (AMB)~\cite{Wittkowski2014AMB}, such that the current in~\eqref{eq_J_quadratic} can be formally written as $\bm J = - \rho M(\varrho) \nabla \psi(\rho)$
where the generalized chemical potential is given by $\psi(\rho) = f'(\rho) - \kappa(\rho)\nabla^2\rho$.
The expressions of the coefficients $f(\rho)$ and $\kappa(\rho)$, as well as details on the derivation, are given in Appendix B.

We deduce that a homogeneous system at density $\br$ is linearly unstable to small wavenumber perturbations whenever $f''(\br) < 0$.
In the limiting cases of large active force and noise, we find that this condition translates to
\begin{subequations}
\begin{align}
\label{eq_MIPS_v0}
\frac{v_0' (2+\alpha)}{v_0(1+\alpha)} \frac{\br}{1+\tfrac{v_D^2}{v_0^2}} & < -1 & (D_v \ll \mu v_0^2) , \\
\label{eq_MIPS_Dv}
\frac{(D_v/\mu)'}{D_v/\mu} \frac{\br}{1+(1+\alpha) \tfrac{v_D^2 \mu}{D_v}} & < -1 & (D_v \gg \mu v_0^2) ,
\end{align}
\end{subequations}
where dependencies of the coefficients in $\rho$ are implicit to lighten the notation, and prime denotes differentiation with respect to $\rho$.
We observe from Eq.~\eqref{eq_MIPS_Dv} that in addition to the standard route, MIPS may also be caused by active fluctuations, as well as cases where the speed relaxation timescale $\mu^{-1}$ depends on the local density.
Moreover, since the instability condition can be fulfilled only if $\mv$ or $\mvsq$ depend on $\rho$ (see Appendix B),
$D_r(\rho)$ by itself cannot lead to MIPS, but will affect the determination of the spinodals in a nontrivial way through the coefficient $\alpha$.
Indeed, taking Eq.~\eqref{eq_MIPS_v0} with $v_D  = 0$ the instability condition becomes
$\br v_0'  / v_0 < -(1+\alpha)/(2+\alpha)$.
For $\alpha \gg 1$, this condition reduces to that of constant speed ABPs~\cite{Cates2015MIPS,Solon2015EPJE},
but smaller values of $\alpha$ in turn lead to a less strict condition such that spinodal decomposition may be observed even if $\br v_0'  / v_0 > -1$.
Finally, when the full dependency of the model parameters on the local particle density is known, 
the MIPS binodals can be computed from $\psi(\rho)$ using the mapping to generalized thermodynamic variables outlined in Ref.~\cite{Solon2018NJP}.

}


We have introduced a general model of active noise and studied its consequences on the free motion of ABPs. 
Our analysis reveals a number of quantitative differences with respect to the widely used constant speed ABPs model,
such as the breakdown of the $\rho_s \sim \langle v \rangle^{-1}$ law for comparable self-propulsion speed and orientational relaxation timescales,
leading under these conditions to possibly counter-intuitive stationary density profiles in activity landscapes.
Moreover, our results illustrate how the interplay of active noise and activity gradients leads to correlations between particles speeds and orientation,
as well as emergent polar order in absence of explicit aligning interactions.
\colrev{Considering a system with quorum sensing interactions, we have shown how active noise modifies the MIPS phase diagram.
In particular, our results suggest that active fluctuations could constitute a mechanistic explanation to the motility induced clustering phenomena observed in systems 
for which the MIPS instability condition for constant speed ABPs does not hold~\cite{Fragkopoulos2021}.}
Considering more complex expressions of the potential $W$ or other types of particle interactions~\cite{Grossmann2012,caprini2022active}, 
it thus appears clear that further consequences of active noise on the dynamics of active matter are expected.

\acknowledgments
This work has received support from the Max Planck School Matter to Life and the MaxSynBio Consortium, which are jointly funded by the Federal Ministry of Education and Research (BMBF) of Germany, and the Max Planck Society.

\vspace{5mm}


\paragraph{Appendix A: the derivation of Eq.~\eqref{eq_J_quadratic}.---}
Here, we provide technical details on the moment expansion and closure procedure leading to the hydrodynamic current~\eqref{eq_J_quadratic} for the density field $\rho$.

Using the definition $\rho \langle \cdot \rangle = \int v^{d-1}{\rm d} {v} {\rm d}\n \, (\cdot) {\cal P}({\bm r},v,\n,t)$ of the velocity moments, 
after some algebra we obtain from Eq.~\eqref{eq_Kramers} and for $k \ge 0$:
\begin{subequations} \label{eq_moments}
\begin{align}
\label{eq:moments_k}
& \partial_t \left(\rho \langle v^k \rangle \right) =
- \nabla \cdot\left[ \rho \langle \hat{\bm n}v^{k+1} \rangle  - D_t\nabla \left(\rho \langle v^k \rangle \right)\right] \\
& \qquad\qquad\qquad\qquad\qquad\qquad\qquad\qquad + \rho\left\langle {\cal G}_k \right\rangle + b_k, \nonumber \\
\label{eq:moments_k_n}
& \partial_t \left(\rho \langle \hat{\bm n} v^k \rangle\right) = 
- \nabla \cdot\left[ \rho \langle {\bm q}v^{k+1} \rangle + \tfrac{\bm I}{d} \rho \langle v^{k+1} \rangle \right.  \\
& \left. - D_t\nabla \left(\rho \langle \hat{\bm n} v^k \rangle\right) \right] 
 - (d-1) D_r  \rho \langle \hat{\bm n} v^k \rangle + \rho\left\langle \hat{\bm n}{\cal G}_k  \right\rangle + \nabla \phi_k, \nonumber
\end{align}
\end{subequations}
where ${\bm q} \equiv \hat{\bm n}\hat{\bm n} - {\bm I}/d$ measures nematic order, and
 ${\cal G}_k \equiv k \left[(k-1)v^{k-2} D_v - v^{k-1}\partial_v W \right] $.
The boundary terms $b_k$ and $\nabla \phi_k$ on the r.h.s.\ of Eqs.~\eqref{eq_moments} 
are obtained after integrating by parts the speed current in Eq.~\eqref{eq_Kramers},
their presence results from the singular behavior of ${\cal P}$ at $v=0$ (see Eq.~\eqref{eq_PH}).
Namely, they are given by
\begin{align*}
b_k & = \lim_{v\to 0} \int{\rm d}\hat{\bm n}\, v^{k+d-1} \left[ \frac{k D_v}{v} - \partial_v W_{\rm eff} - D_v \partial_v \right] {\cal P} , \\
\nabla \phi_k & = \lim_{v\to 0} \int{\rm d}\hat{\bm n}\,  \hat{\bm n} v^{k+d-1} \left[ \frac{k D_v}{v} - \partial_v W_{\rm eff} - D_v \partial_v \right] {\cal P}.
\end{align*}
To evaluate them, we show in~\cite{Note1} that the distribution ${\cal P}(\bm r,v,\n,t)$ can be written perturbatively as
\begin{equation*}
{\cal P}(\bm r,v,\n,t) \simeq \rho(\bm r,t)\frac{{\cal P}_{\rm H}(\bm r,v)}{{\cal S}_d}\left[ 1 + Y(\bm r, v,\hat{\bm n})  \right],
\end{equation*}
where 
${\cal S}_d$ denotes the surface of the unit ($d-1$)-sphere, 
${\cal P}_{\rm H}$ is given by~\eqref{eq_PH} with space-dependent $W$ and $D_v$, 
while the ${\cal O}(\n\cdot\nabla)$ function $Y$ is unknown but satisfies in the regime $D_v \gg W$ relevant for the boundary terms:
\begin{equation*}
Y(\bm r, v,\hat{\bm n})  \underset{W\gg D_v}{\simeq} - v\tau_r (\hat{\bm n}\cdot \nabla)  \ln\left[{\cal P}_{\rm H}(\bm r,v)\right]
\end{equation*}
Using these expressions, we therefore obtain
\begin{equation}
b_k =  D_v \rho \gamma \, \delta_{k,1} , \qquad \nabla \phi_k = \frac{D_v \tau_r \delta_{k,0}}{d}\nabla \left( \rho \gamma \right),
\end{equation}
where $\gamma(\bm r) \equiv {\cal P}_{\rm H}(\bm r,0) = e^{-W(\bm r,0)/D_v(\bm r)}Z^{-1}(\bm r)$ and $\delta_{i,j}$ is the Kronecker delta symbol.

Considering now $ W({\bm r},v) = \tfrac{\mu(\bm r)}{2} \left[ v - v_0(\bm r)\right]^2 $, we express from Eq.~\eqref{eq:moments_k_n} the first two polarity moments as
\begin{subequations} \label{eq_polar_moments}
\begin{align}
 \label{eq_n}
\rho \langle \hat{\bm n} \rangle &= -\frac{\tau_r}{d}\nabla (\rho  \langle v \rangle) + \frac{D_v\tau_r^2}{d}\nabla \left(\rho \gamma \right) , \\
 \label{eq_vn}
\rho \langle v \hat{\bm n} \rangle &= -\frac{\tau_r}{d(1+\alpha)}\nabla \left(\rho  \langle v^2 \rangle \right)+ \frac{\alpha v_0}{1+\alpha} \rho \langle \hat{\bm n} \rangle ,
\end{align} 
\end{subequations}
where $\alpha \equiv \mu \tau_r$ and we have kept only contributions up to ${\cal O}(\nabla)$.
In particular, the contribution from the nematic order parameter $\bm q$ was discarded as it is ${\cal O}(\nabla^2)$~\cite{Note1}.
At first order in gradients the self-propulsion speed moments moreover solve $\langle {\cal G}_k \rangle + b_k = 0$,
such that $\langle v^k \rangle = \langle v^k \rangle_{\rm H}$ where the $\rm H$ subscript indicates that the average is taken w.r.t.\ the distribution $\calP_{\rm H}(\bm r,v)$.
Noting that the density satisfies from~\eqref{eq:moments_k} with $k=0$ 
\begin{equation*}
\partial_t \rho = -\nabla \cdot \left[ \rho \langle v\hat{\bm n} \rangle  - D_t\nabla \rho \right],
\end{equation*}
and combining Eqs.~\eqref{eq_speed_moments_quadratic} and~\eqref{eq_polar_moments} finally gives the hydrodynamic current~\eqref{eq_J_quadratic}.
Note that the boundary term $\sim \tau_r^2 D_v v_0 \nabla \left(\rho \gamma\right)$ coming from~\eqref{eq_n} has been discarded as its contribution is generally subleading.

\colrev{
\paragraph{Appendix B: the mapping to AMB.---}

Here, we present the derivation of the nonlocal currents arising when the interactions between the particles can be treated as effective quorum sensing interactions,
leading to a dependence in the particle density of the hydrodynamic equation coefficients.
Let us therefore consider the hydrodynamic current~\eqref{eq_J_quadratic}
and assume that all coefficients take a functional dependency in the local density field such that
\begin{equation} \label{eq_hydro_coeffs_SM}
\langle v^2 \rangle(\varrho), \; \langle v \rangle(\varrho), \; v_0(\varrho), \; \tau_r(\varrho), \; \alpha(\varrho), \; D_t(\varrho) ,
\end{equation}
where $\varrho \equiv \int\rmd^d\bm r'\, K(|\bm r' - \bm r|)\rho(\bm r')$ and $K$ is a short-range interaction kernel normalized such that $\int\rmd^d\bm r'\, K(r') = 1$.
We now define
\begin{equation}
   M(\varrho) \equiv D_t(\varrho) + \frac{\tau_r(\varrho)}{d(1+\alpha(\varrho))} \left[\langle v^2 \rangle(\varrho) + \alpha(\varrho) v_0(\varrho) \langle v \rangle(\varrho) \right], 
\end{equation}
as an effective mobility for the dynamics. 
Note that this choice is arbitrary so long as $M(\varrho)>0$.
The current in Eq.~\eqref{eq_J_quadratic} is then written as
\begin{align} \label{eq_SM_Jintermediate}
    \bm J = & -\rho M(\varrho) \Big[ 
    \nabla \ln(\rho) +  \\
    & \left. \frac{\nabla [\langle v^2 \rangle(\varrho)] + \alpha(\varrho)v_0(\varrho)\nabla[\langle v \rangle(\varrho)] }{(1+\alpha(\varrho))v_D^2(\varrho) + \langle v^2 \rangle(\varrho) + \alpha(\varrho) v_0(\varrho) \langle v \rangle(\varrho)} \nonumber
    \right],
\end{align}
where $v_D^2 \equiv d D_t \tau_r^{-1}$.
Assuming that $\rho$ does not vary much over the scale of the quorum sensing interaction, we expand for any function $\phi(\varrho)$ up to second order in gradient:
\begin{equation*}
    \phi(\varrho) \simeq \phi\left( \rho + \ell^2 \nabla^2\rho \right) 
    \simeq \phi(\rho) + \ell^2 \phi'(\rho) \nabla^2\rho,
\end{equation*}
where $\ell^2 \equiv  \tfrac{1}{2d}\int \rmd^d \bm r\, K(r) r^2$ and prime denotes derivative w.r.t.\ $\rho$.
Expanding all coefficients in the fraction on the r.h.s.\ of~\eqref{eq_SM_Jintermediate}, we find after some algebra that the current can be recast into the compact form $\bm J = - \rho M(\varrho) \nabla \psi(\rho)$,
where the effective chemical potential $\psi(\rho) \equiv f'(\rho) - \kappa(\rho)\nabla^2\rho$, while the generalized free energy and surface tension are defined as
\begin{align*}
f'(\rho) & \equiv \ln(\rho) + \int^\rho \rmd y \, \frac{\langle v^2 \rangle' + \alpha v_0 \langle v \rangle'}{(1+\alpha)v_D^2 + \langle v^2 \rangle + \alpha v_0 \langle v \rangle}, \\
\kappa(\rho) & \equiv -\ell^2\frac{\langle v^2 \rangle' + \alpha v_0 \langle v \rangle'}{(1+\alpha)v_D^2 + \langle v^2 \rangle + \alpha v_0 \langle v \rangle},
\end{align*}
where the $\rho$-dependency of the coefficients is kept implicit to lighten notations.
}

\bibliography{bibliograph}

\begin{thebibliography}{60}%
\makeatletter
\providecommand \@ifxundefined [1]{%
 \@ifx{#1\undefined}
}%
\providecommand \@ifnum [1]{%
 \ifnum #1\expandafter \@firstoftwo
 \else \expandafter \@secondoftwo
 \fi
}%
\providecommand \@ifx [1]{%
 \ifx #1\expandafter \@firstoftwo
 \else \expandafter \@secondoftwo
 \fi
}%
\providecommand \natexlab [1]{#1}%
\providecommand \enquote  [1]{``#1''}%
\providecommand \bibnamefont  [1]{#1}%
\providecommand \bibfnamefont [1]{#1}%
\providecommand \citenamefont [1]{#1}%
\providecommand \href@noop [0]{\@secondoftwo}%
\providecommand \href [0]{\begingroup \@sanitize@url \@href}%
\providecommand \@href[1]{\@@startlink{#1}\@@href}%
\providecommand \@@href[1]{\endgroup#1\@@endlink}%
\providecommand \@sanitize@url [0]{\catcode `\\12\catcode `\$12\catcode
  `\&12\catcode `\#12\catcode `\^12\catcode `\_12\catcode `\%12\relax}%
\providecommand \@@startlink[1]{}%
\providecommand \@@endlink[0]{}%
\providecommand \url  [0]{\begingroup\@sanitize@url \@url }%
\providecommand \@url [1]{\endgroup\@href {#1}{\urlprefix }}%
\providecommand \urlprefix  [0]{URL }%
\providecommand \Eprint [0]{\href }%
\providecommand \doibase [0]{https://doi.org/}%
\providecommand \selectlanguage [0]{\@gobble}%
\providecommand \bibinfo  [0]{\@secondoftwo}%
\providecommand \bibfield  [0]{\@secondoftwo}%
\providecommand \translation [1]{[#1]}%
\providecommand \BibitemOpen [0]{}%
\providecommand \bibitemStop [0]{}%
\providecommand \bibitemNoStop [0]{.\EOS\space}%
\providecommand \EOS [0]{\spacefactor3000\relax}%
\providecommand \BibitemShut  [1]{\csname bibitem#1\endcsname}%
\let\auto@bib@innerbib\@empty
\bibitem [{\citenamefont {Prost}\ \emph {et~al.}(2015)\citenamefont {Prost},
  \citenamefont {J{\"u}licher},\ and\ \citenamefont
  {Joanny}}]{Prost2015NatPhys}%
  \BibitemOpen
  \bibfield  {author} {\bibinfo {author} {\bibfnamefont {J.}~\bibnamefont
  {Prost}}, \bibinfo {author} {\bibfnamefont {F.}~\bibnamefont
  {J{\"u}licher}},\ and\ \bibinfo {author} {\bibfnamefont {J.-F.}\ \bibnamefont
  {Joanny}},\ }\bibfield  {title} {\bibinfo {title} {Active gel physics},\
  }\href {https://doi.org/10.1038/nphys3224} {\bibfield  {journal} {\bibinfo
  {journal} {Nat. Phys.}\ }\textbf {\bibinfo {volume} {11}},\ \bibinfo {pages}
  {111} (\bibinfo {year} {2015})}\BibitemShut {NoStop}%
\bibitem [{\citenamefont {Doostmohammadi}\ \emph {et~al.}(2018)\citenamefont
  {Doostmohammadi}, \citenamefont {Ign{\'e}s-Mullol}, \citenamefont {Yeomans},\
  and\ \citenamefont {Sagu{\'e}s}}]{Doostmohammadi2018NatComm}%
  \BibitemOpen
  \bibfield  {author} {\bibinfo {author} {\bibfnamefont {A.}~\bibnamefont
  {Doostmohammadi}}, \bibinfo {author} {\bibfnamefont {J.}~\bibnamefont
  {Ign{\'e}s-Mullol}}, \bibinfo {author} {\bibfnamefont {J.~M.}\ \bibnamefont
  {Yeomans}},\ and\ \bibinfo {author} {\bibfnamefont {F.}~\bibnamefont
  {Sagu{\'e}s}},\ }\bibfield  {title} {\bibinfo {title} {Active nematics},\
  }\href {https://doi.org/10.1038/s41467-018-05666-8} {\bibfield  {journal}
  {\bibinfo  {journal} {Nat. Commun.}\ }\textbf {\bibinfo {volume} {9}},\
  \bibinfo {pages} {3246} (\bibinfo {year} {2018})}\BibitemShut {NoStop}%
\bibitem [{\citenamefont {Golestanian}(2022)}]{Golestanian2019phoretic}%
  \BibitemOpen
  \bibfield  {author} {\bibinfo {author} {\bibfnamefont {R.}~\bibnamefont
  {Golestanian}},\ }\bibfield  {title} {\bibinfo {title} {{Phoretic Active
  Matter}},\ }in\ \href {https://doi.org/10.1093/oso/9780192858313.003.0008}
  {\emph {\bibinfo {booktitle} {{Active Matter and Nonequilibrium Statistical
  Physics: Lecture Notes of the Les Houches Summer School: Volume 112,
  September 2018}}}}\ (\bibinfo  {publisher} {Oxford University Press},\
  \bibinfo {year} {2022})\BibitemShut {NoStop}%
\bibitem [{\citenamefont {Soto}\ and\ \citenamefont
  {Golestanian}(2014)}]{Soto:2014}%
  \BibitemOpen
  \bibfield  {author} {\bibinfo {author} {\bibfnamefont {R.}~\bibnamefont
  {Soto}}\ and\ \bibinfo {author} {\bibfnamefont {R.}~\bibnamefont
  {Golestanian}},\ }\bibfield  {title} {\bibinfo {title} {Self-assembly of
  catalytically active colloidal molecules: Tailoring activity through surface
  chemistry},\ }\href {https://doi.org/10.1103/PhysRevLett.112.068301}
  {\bibfield  {journal} {\bibinfo  {journal} {Phys. Rev. Lett.}\ }\textbf
  {\bibinfo {volume} {112}},\ \bibinfo {pages} {068301} (\bibinfo {year}
  {2014})}\BibitemShut {NoStop}%
\bibitem [{\citenamefont {Saha}\ \emph {et~al.}(2019)\citenamefont {Saha},
  \citenamefont {Ramaswamy},\ and\ \citenamefont {Golestanian}}]{Saha2019}%
  \BibitemOpen
  \bibfield  {author} {\bibinfo {author} {\bibfnamefont {S.}~\bibnamefont
  {Saha}}, \bibinfo {author} {\bibfnamefont {S.}~\bibnamefont {Ramaswamy}},\
  and\ \bibinfo {author} {\bibfnamefont {R.}~\bibnamefont {Golestanian}},\
  }\bibfield  {title} {\bibinfo {title} {Pairing, waltzing and scattering of
  chemotactic active colloids},\ }\href
  {https://doi.org/10.1088/1367-2630/ab20fd} {\bibfield  {journal} {\bibinfo
  {journal} {New Journal of Physics}\ }\textbf {\bibinfo {volume} {21}},\
  \bibinfo {pages} {063006} (\bibinfo {year} {2019})}\BibitemShut {NoStop}%
\bibitem [{\citenamefont {Saha}\ \emph {et~al.}(2020)\citenamefont {Saha},
  \citenamefont {Agudo-Canalejo},\ and\ \citenamefont
  {Golestanian}}]{Saha2020PRX}%
  \BibitemOpen
  \bibfield  {author} {\bibinfo {author} {\bibfnamefont {S.}~\bibnamefont
  {Saha}}, \bibinfo {author} {\bibfnamefont {J.}~\bibnamefont
  {Agudo-Canalejo}},\ and\ \bibinfo {author} {\bibfnamefont {R.}~\bibnamefont
  {Golestanian}},\ }\bibfield  {title} {\bibinfo {title} {Scalar active
  mixtures: The nonreciprocal cahn-hilliard model},\ }\href
  {https://doi.org/10.1103/PhysRevX.10.041009} {\bibfield  {journal} {\bibinfo
  {journal} {Phys. Rev. X}\ }\textbf {\bibinfo {volume} {10}},\ \bibinfo
  {pages} {041009} (\bibinfo {year} {2020})}\BibitemShut {NoStop}%
\bibitem [{\citenamefont {You}\ \emph {et~al.}(2020)\citenamefont {You},
  \citenamefont {Baskaran},\ and\ \citenamefont {Marchetti}}]{You2020PNAS}%
  \BibitemOpen
  \bibfield  {author} {\bibinfo {author} {\bibfnamefont {Z.}~\bibnamefont
  {You}}, \bibinfo {author} {\bibfnamefont {A.}~\bibnamefont {Baskaran}},\ and\
  \bibinfo {author} {\bibfnamefont {M.~C.}\ \bibnamefont {Marchetti}},\
  }\bibfield  {title} {\bibinfo {title} {Nonreciprocity as a generic route to
  traveling states},\ }\href {https://doi.org/10.1073/pnas.2010318117}
  {\bibfield  {journal} {\bibinfo  {journal} {Proc. Natl. Acad. Sci. U.S.A.}\
  }\textbf {\bibinfo {volume} {117}},\ \bibinfo {pages} {19767} (\bibinfo
  {year} {2020})}\BibitemShut {NoStop}%
\bibitem [{\citenamefont {Dadhichi}\ \emph {et~al.}(2020)\citenamefont
  {Dadhichi}, \citenamefont {Kethapelli}, \citenamefont {Chajwa}, \citenamefont
  {Ramaswamy},\ and\ \citenamefont {Maitra}}]{DadhichiPRE2020}%
  \BibitemOpen
  \bibfield  {author} {\bibinfo {author} {\bibfnamefont {L.~P.}\ \bibnamefont
  {Dadhichi}}, \bibinfo {author} {\bibfnamefont {J.}~\bibnamefont
  {Kethapelli}}, \bibinfo {author} {\bibfnamefont {R.}~\bibnamefont {Chajwa}},
  \bibinfo {author} {\bibfnamefont {S.}~\bibnamefont {Ramaswamy}},\ and\
  \bibinfo {author} {\bibfnamefont {A.}~\bibnamefont {Maitra}},\ }\bibfield
  {title} {\bibinfo {title} {Nonmutual torques and the unimportance of motility
  for long-range order in two-dimensional flocks},\ }\href
  {https://doi.org/10.1103/PhysRevE.101.052601} {\bibfield  {journal} {\bibinfo
   {journal} {Phys. Rev. E}\ }\textbf {\bibinfo {volume} {101}},\ \bibinfo
  {pages} {052601} (\bibinfo {year} {2020})}\BibitemShut {NoStop}%
\bibitem [{\citenamefont {Fruchart}\ \emph {et~al.}(2021)\citenamefont
  {Fruchart}, \citenamefont {Hanai}, \citenamefont {Littlewood},\ and\
  \citenamefont {Vitelli}}]{Fruchart2021Nat}%
  \BibitemOpen
  \bibfield  {author} {\bibinfo {author} {\bibfnamefont {M.}~\bibnamefont
  {Fruchart}}, \bibinfo {author} {\bibfnamefont {R.}~\bibnamefont {Hanai}},
  \bibinfo {author} {\bibfnamefont {P.~B.}\ \bibnamefont {Littlewood}},\ and\
  \bibinfo {author} {\bibfnamefont {V.}~\bibnamefont {Vitelli}},\ }\bibfield
  {title} {\bibinfo {title} {Non-reciprocal phase transitions},\ }\href
  {https://doi.org/10.1038/s41586-021-03375-9} {\bibfield  {journal} {\bibinfo
  {journal} {Nature}\ }\textbf {\bibinfo {volume} {592}},\ \bibinfo {pages}
  {363} (\bibinfo {year} {2021})}\BibitemShut {NoStop}%
\bibitem [{\citenamefont {Dell'Arciprete}\ \emph {et~al.}(2018)\citenamefont
  {Dell'Arciprete}, \citenamefont {Blow}, \citenamefont {Brown}, \citenamefont
  {Farrell}, \citenamefont {Lintuvuori}, \citenamefont {McVey}, \citenamefont
  {Marenduzzo},\ and\ \citenamefont {Poon}}]{DellArciprete2018NatComm}%
  \BibitemOpen
  \bibfield  {author} {\bibinfo {author} {\bibfnamefont {D.}~\bibnamefont
  {Dell'Arciprete}}, \bibinfo {author} {\bibfnamefont {M.~L.}\ \bibnamefont
  {Blow}}, \bibinfo {author} {\bibfnamefont {A.~T.}\ \bibnamefont {Brown}},
  \bibinfo {author} {\bibfnamefont {F.~D.~C.}\ \bibnamefont {Farrell}},
  \bibinfo {author} {\bibfnamefont {J.~S.}\ \bibnamefont {Lintuvuori}},
  \bibinfo {author} {\bibfnamefont {A.~F.}\ \bibnamefont {McVey}}, \bibinfo
  {author} {\bibfnamefont {D.}~\bibnamefont {Marenduzzo}},\ and\ \bibinfo
  {author} {\bibfnamefont {W.~C.~K.}\ \bibnamefont {Poon}},\ }\bibfield
  {title} {\bibinfo {title} {A growing bacterial colony in two dimensions as an
  active nematic},\ }\href {https://doi.org/10.1038/s41467-018-06370-3}
  {\bibfield  {journal} {\bibinfo  {journal} {Nat. Commun.}\ }\textbf {\bibinfo
  {volume} {9}},\ \bibinfo {pages} {4190} (\bibinfo {year} {2018})}\BibitemShut
  {NoStop}%
\bibitem [{\citenamefont {You}\ \emph {et~al.}(2018)\citenamefont {You},
  \citenamefont {Pearce}, \citenamefont {Sengupta},\ and\ \citenamefont
  {Giomi}}]{You2018PRX}%
  \BibitemOpen
  \bibfield  {author} {\bibinfo {author} {\bibfnamefont {Z.}~\bibnamefont
  {You}}, \bibinfo {author} {\bibfnamefont {D.~J.~G.}\ \bibnamefont {Pearce}},
  \bibinfo {author} {\bibfnamefont {A.}~\bibnamefont {Sengupta}},\ and\
  \bibinfo {author} {\bibfnamefont {L.}~\bibnamefont {Giomi}},\ }\bibfield
  {title} {\bibinfo {title} {Geometry and mechanics of microdomains in growing
  bacterial colonies},\ }\href {https://doi.org/10.1103/PhysRevX.8.031065}
  {\bibfield  {journal} {\bibinfo  {journal} {Phys. Rev. X}\ }\textbf {\bibinfo
  {volume} {8}},\ \bibinfo {pages} {031065} (\bibinfo {year}
  {2018})}\BibitemShut {NoStop}%
\bibitem [{\citenamefont {Hartmann}\ \emph {et~al.}(2019)\citenamefont
  {Hartmann}, \citenamefont {Singh}, \citenamefont {Pearce}, \citenamefont
  {Mok}, \citenamefont {Song}, \citenamefont {D{\'\i}az-Pascual}, \citenamefont
  {Dunkel},\ and\ \citenamefont {Drescher}}]{Hartmann2019NatPhys}%
  \BibitemOpen
  \bibfield  {author} {\bibinfo {author} {\bibfnamefont {R.}~\bibnamefont
  {Hartmann}}, \bibinfo {author} {\bibfnamefont {P.~K.}\ \bibnamefont {Singh}},
  \bibinfo {author} {\bibfnamefont {P.}~\bibnamefont {Pearce}}, \bibinfo
  {author} {\bibfnamefont {R.}~\bibnamefont {Mok}}, \bibinfo {author}
  {\bibfnamefont {B.}~\bibnamefont {Song}}, \bibinfo {author} {\bibfnamefont
  {F.}~\bibnamefont {D{\'\i}az-Pascual}}, \bibinfo {author} {\bibfnamefont
  {J.}~\bibnamefont {Dunkel}},\ and\ \bibinfo {author} {\bibfnamefont
  {K.}~\bibnamefont {Drescher}},\ }\bibfield  {title} {\bibinfo {title}
  {Emergence of three-dimensional order and structure in growing biofilms},\
  }\href {https://doi.org/10.1038/s41567-018-0356-9} {\bibfield  {journal}
  {\bibinfo  {journal} {Nat. Phys.}\ }\textbf {\bibinfo {volume} {15}},\
  \bibinfo {pages} {251} (\bibinfo {year} {2019})}\BibitemShut {NoStop}%
\bibitem [{\citenamefont {Romanczuk}\ \emph {et~al.}(2012)\citenamefont
  {Romanczuk}, \citenamefont {B{\"a}r}, \citenamefont {Ebeling}, \citenamefont
  {Lindner},\ and\ \citenamefont {Schimansky-Geier}}]{ABPreview2012}%
  \BibitemOpen
  \bibfield  {author} {\bibinfo {author} {\bibfnamefont {P.}~\bibnamefont
  {Romanczuk}}, \bibinfo {author} {\bibfnamefont {M.}~\bibnamefont {B{\"a}r}},
  \bibinfo {author} {\bibfnamefont {W.}~\bibnamefont {Ebeling}}, \bibinfo
  {author} {\bibfnamefont {B.}~\bibnamefont {Lindner}},\ and\ \bibinfo {author}
  {\bibfnamefont {L.}~\bibnamefont {Schimansky-Geier}},\ }\bibfield  {title}
  {\bibinfo {title} {Active brownian particles},\ }\href
  {https://doi.org/10.1140/epjst/e2012-01529-y} {\bibfield  {journal} {\bibinfo
   {journal} {Eur. Phys. J. Spec. Top.}\ }\textbf {\bibinfo {volume} {202}},\
  \bibinfo {pages} {1} (\bibinfo {year} {2012})}\BibitemShut {NoStop}%
\bibitem [{\citenamefont {Marchetti}\ \emph {et~al.}(2013)\citenamefont
  {Marchetti}, \citenamefont {Joanny}, \citenamefont {Ramaswamy}, \citenamefont
  {Liverpool}, \citenamefont {Prost}, \citenamefont {Rao},\ and\ \citenamefont
  {Simha}}]{MarchettiRMP2013}%
  \BibitemOpen
  \bibfield  {author} {\bibinfo {author} {\bibfnamefont {M.~C.}\ \bibnamefont
  {Marchetti}}, \bibinfo {author} {\bibfnamefont {J.~F.}\ \bibnamefont
  {Joanny}}, \bibinfo {author} {\bibfnamefont {S.}~\bibnamefont {Ramaswamy}},
  \bibinfo {author} {\bibfnamefont {T.~B.}\ \bibnamefont {Liverpool}}, \bibinfo
  {author} {\bibfnamefont {J.}~\bibnamefont {Prost}}, \bibinfo {author}
  {\bibfnamefont {M.}~\bibnamefont {Rao}},\ and\ \bibinfo {author}
  {\bibfnamefont {R.~A.}\ \bibnamefont {Simha}},\ }\bibfield  {title} {\bibinfo
  {title} {Hydrodynamics of soft active matter},\ }\href
  {https://doi.org/10.1103/RevModPhys.85.1143} {\bibfield  {journal} {\bibinfo
  {journal} {Rev. Mod. Phys.}\ }\textbf {\bibinfo {volume} {85}},\ \bibinfo
  {pages} {1143} (\bibinfo {year} {2013})}\BibitemShut {NoStop}%
\bibitem [{\citenamefont {Cates}\ and\ \citenamefont
  {Tailleur}(2015)}]{Cates2015MIPS}%
  \BibitemOpen
  \bibfield  {author} {\bibinfo {author} {\bibfnamefont {M.~E.}\ \bibnamefont
  {Cates}}\ and\ \bibinfo {author} {\bibfnamefont {J.}~\bibnamefont
  {Tailleur}},\ }\bibfield  {title} {\bibinfo {title} {Motility-induced phase
  separation},\ }\href
  {https://doi.org/10.1146/annurev-conmatphys-031214-014710} {\bibfield
  {journal} {\bibinfo  {journal} {Ann. Rev. Cond. Mat. Phys.}\ }\textbf
  {\bibinfo {volume} {6}},\ \bibinfo {pages} {219} (\bibinfo {year}
  {2015})}\BibitemShut {NoStop}%
\bibitem [{\citenamefont {Gompper}\ and\ \citenamefont {\textit{et
  al.}}(2020)}]{Gompper2020IOP}%
  \BibitemOpen
  \bibfield  {author} {\bibinfo {author} {\bibfnamefont {G.}~\bibnamefont
  {Gompper}}\ and\ \bibinfo {author} {\bibnamefont {\textit{et al.}}},\
  }\bibfield  {title} {\bibinfo {title} {The 2020 motile active matter
  roadmap},\ }\href {https://doi.org/10.1088/1361-648x/ab6348} {\bibfield
  {journal} {\bibinfo  {journal} {J. Phys.: Condens. Matter}\ }\textbf
  {\bibinfo {volume} {32}},\ \bibinfo {pages} {193001} (\bibinfo {year}
  {2020})}\BibitemShut {NoStop}%
\bibitem [{\citenamefont {Chat{\'e}}(2020)}]{ChateDADAM2020}%
  \BibitemOpen
  \bibfield  {author} {\bibinfo {author} {\bibfnamefont {H.}~\bibnamefont
  {Chat{\'e}}},\ }\bibfield  {title} {\bibinfo {title} {{Dry Aligning Dilute
  Active Matter}},\ }\href
  {https://doi.org/10.1146/annurev-conmatphys-031119-050752} {\bibfield
  {journal} {\bibinfo  {journal} {Annu. Rev. Condens. Matter Phys.}\ }\textbf
  {\bibinfo {volume} {11}},\ \bibinfo {pages} {189} (\bibinfo {year}
  {2020})}\BibitemShut {NoStop}%
\bibitem [{\citenamefont {Gao}\ and\ \citenamefont {Wang}(2014)}]{Gao2014ACS}%
  \BibitemOpen
  \bibfield  {author} {\bibinfo {author} {\bibfnamefont {W.}~\bibnamefont
  {Gao}}\ and\ \bibinfo {author} {\bibfnamefont {J.}~\bibnamefont {Wang}},\
  }\bibfield  {title} {\bibinfo {title} {The environmental impact of
  micro/nanomachines: A review},\ }\bibfield  {booktitle} {\emph {\bibinfo
  {booktitle} {ACS Nano}},\ }\href {https://doi.org/10.1021/nn500077a}
  {\bibfield  {journal} {\bibinfo  {journal} {ACS Nano}\ }\textbf {\bibinfo
  {volume} {8}},\ \bibinfo {pages} {3170} (\bibinfo {year} {2014})}\BibitemShut
  {NoStop}%
\bibitem [{\citenamefont {Ghosh}\ \emph {et~al.}(2020)\citenamefont {Ghosh},
  \citenamefont {Xu}, \citenamefont {Gupta},\ and\ \citenamefont
  {Gracias}}]{Ghosh2020Nano}%
  \BibitemOpen
  \bibfield  {author} {\bibinfo {author} {\bibfnamefont {A.}~\bibnamefont
  {Ghosh}}, \bibinfo {author} {\bibfnamefont {W.}~\bibnamefont {Xu}}, \bibinfo
  {author} {\bibfnamefont {N.}~\bibnamefont {Gupta}},\ and\ \bibinfo {author}
  {\bibfnamefont {D.~H.}\ \bibnamefont {Gracias}},\ }\bibfield  {title}
  {\bibinfo {title} {Active matter therapeutics},\ }\href
  {https://doi.org/10.1016/j.nantod.2019.100836} {\bibfield  {journal}
  {\bibinfo  {journal} {Nano today}\ }\textbf {\bibinfo {volume} {31}},\
  \bibinfo {pages} {100836} (\bibinfo {year} {2020})}\BibitemShut {NoStop}%
\bibitem [{\citenamefont {Lozano}\ \emph {et~al.}(2016)\citenamefont {Lozano},
  \citenamefont {ten Hagen}, \citenamefont {L{\"o}wen},\ and\ \citenamefont
  {Bechinger}}]{Lozano2016NatComm}%
  \BibitemOpen
  \bibfield  {author} {\bibinfo {author} {\bibfnamefont {C.}~\bibnamefont
  {Lozano}}, \bibinfo {author} {\bibfnamefont {B.}~\bibnamefont {ten Hagen}},
  \bibinfo {author} {\bibfnamefont {H.}~\bibnamefont {L{\"o}wen}},\ and\
  \bibinfo {author} {\bibfnamefont {C.}~\bibnamefont {Bechinger}},\ }\bibfield
  {title} {\bibinfo {title} {Phototaxis of synthetic microswimmers in optical
  landscapes},\ }\href {https://doi.org/10.1038/ncomms12828} {\bibfield
  {journal} {\bibinfo  {journal} {Nat. Commun.}\ }\textbf {\bibinfo {volume}
  {7}},\ \bibinfo {pages} {12828} (\bibinfo {year} {2016})}\BibitemShut
  {NoStop}%
\bibitem [{\citenamefont {Arlt}\ \emph {et~al.}(2018)\citenamefont {Arlt},
  \citenamefont {Martinez}, \citenamefont {Dawson}, \citenamefont {Pilizota},\
  and\ \citenamefont {Poon}}]{Arlt2018NatComm}%
  \BibitemOpen
  \bibfield  {author} {\bibinfo {author} {\bibfnamefont {J.}~\bibnamefont
  {Arlt}}, \bibinfo {author} {\bibfnamefont {V.~A.}\ \bibnamefont {Martinez}},
  \bibinfo {author} {\bibfnamefont {A.}~\bibnamefont {Dawson}}, \bibinfo
  {author} {\bibfnamefont {T.}~\bibnamefont {Pilizota}},\ and\ \bibinfo
  {author} {\bibfnamefont {W.~C.~K.}\ \bibnamefont {Poon}},\ }\bibfield
  {title} {\bibinfo {title} {Painting with light-powered bacteria},\ }\href
  {https://doi.org/10.1038/s41467-018-03161-8} {\bibfield  {journal} {\bibinfo
  {journal} {Nat. Commun.}\ }\textbf {\bibinfo {volume} {9}},\ \bibinfo {pages}
  {768} (\bibinfo {year} {2018})}\BibitemShut {NoStop}%
\bibitem [{\citenamefont {Frangipane}\ \emph {et~al.}(2018)\citenamefont
  {Frangipane}, \citenamefont {Dell'Arciprete}, \citenamefont {Petracchini},
  \citenamefont {Maggi}, \citenamefont {Saglimbeni}, \citenamefont {Bianchi},
  \citenamefont {Vizsnyiczai}, \citenamefont {Bernardini},\ and\ \citenamefont
  {Di~Leonardo}}]{frangipane2018Elife}%
  \BibitemOpen
  \bibfield  {author} {\bibinfo {author} {\bibfnamefont {G.}~\bibnamefont
  {Frangipane}}, \bibinfo {author} {\bibfnamefont {D.}~\bibnamefont
  {Dell'Arciprete}}, \bibinfo {author} {\bibfnamefont {S.}~\bibnamefont
  {Petracchini}}, \bibinfo {author} {\bibfnamefont {C.}~\bibnamefont {Maggi}},
  \bibinfo {author} {\bibfnamefont {F.}~\bibnamefont {Saglimbeni}}, \bibinfo
  {author} {\bibfnamefont {S.}~\bibnamefont {Bianchi}}, \bibinfo {author}
  {\bibfnamefont {G.}~\bibnamefont {Vizsnyiczai}}, \bibinfo {author}
  {\bibfnamefont {M.~L.}\ \bibnamefont {Bernardini}},\ and\ \bibinfo {author}
  {\bibfnamefont {R.}~\bibnamefont {Di~Leonardo}},\ }\bibfield  {title}
  {\bibinfo {title} {Dynamic density shaping of photokinetic e. coli},\ }\href
  {https://elifesciences.org/articles/36608} {\bibfield  {journal} {\bibinfo
  {journal} {Elife}\ }\textbf {\bibinfo {volume} {7}},\ \bibinfo {pages}
  {e36608} (\bibinfo {year} {2018})}\BibitemShut {NoStop}%
\bibitem [{\citenamefont {S\"oker}\ \emph {et~al.}(2021)\citenamefont
  {S\"oker}, \citenamefont {Auschra}, \citenamefont {Holubec}, \citenamefont
  {Kroy},\ and\ \citenamefont {Cichos}}]{Soker2021PRL}%
  \BibitemOpen
  \bibfield  {author} {\bibinfo {author} {\bibfnamefont {N.~A.}\ \bibnamefont
  {S\"oker}}, \bibinfo {author} {\bibfnamefont {S.}~\bibnamefont {Auschra}},
  \bibinfo {author} {\bibfnamefont {V.}~\bibnamefont {Holubec}}, \bibinfo
  {author} {\bibfnamefont {K.}~\bibnamefont {Kroy}},\ and\ \bibinfo {author}
  {\bibfnamefont {F.}~\bibnamefont {Cichos}},\ }\bibfield  {title} {\bibinfo
  {title} {How activity landscapes polarize microswimmers without alignment
  forces},\ }\href {https://doi.org/10.1103/PhysRevLett.126.228001} {\bibfield
  {journal} {\bibinfo  {journal} {Phys. Rev. Lett.}\ }\textbf {\bibinfo
  {volume} {126}},\ \bibinfo {pages} {228001} (\bibinfo {year}
  {2021})}\BibitemShut {NoStop}%
\bibitem [{\citenamefont {B{\"a}uerle}\ \emph {et~al.}(2018)\citenamefont
  {B{\"a}uerle}, \citenamefont {Fischer}, \citenamefont {Speck},\ and\
  \citenamefont {Bechinger}}]{Bauerle2018NatComm}%
  \BibitemOpen
  \bibfield  {author} {\bibinfo {author} {\bibfnamefont {T.}~\bibnamefont
  {B{\"a}uerle}}, \bibinfo {author} {\bibfnamefont {A.}~\bibnamefont
  {Fischer}}, \bibinfo {author} {\bibfnamefont {T.}~\bibnamefont {Speck}},\
  and\ \bibinfo {author} {\bibfnamefont {C.}~\bibnamefont {Bechinger}},\
  }\bibfield  {title} {\bibinfo {title} {Self-organization of active particles
  by quorum sensing rules},\ }\href
  {https://doi.org/10.1038/s41467-018-05675-7} {\bibfield  {journal} {\bibinfo
  {journal} {Nat. Commun.}\ }\textbf {\bibinfo {volume} {9}},\ \bibinfo {pages}
  {3232} (\bibinfo {year} {2018})}\BibitemShut {NoStop}%
\bibitem [{\citenamefont {Lavergne}\ \emph {et~al.}(2019)\citenamefont
  {Lavergne}, \citenamefont {Wendehenne}, \citenamefont {B{\"a}uerle},\ and\
  \citenamefont {Bechinger}}]{Lavergne2019}%
  \BibitemOpen
  \bibfield  {author} {\bibinfo {author} {\bibfnamefont {F.~A.}\ \bibnamefont
  {Lavergne}}, \bibinfo {author} {\bibfnamefont {H.}~\bibnamefont
  {Wendehenne}}, \bibinfo {author} {\bibfnamefont {T.}~\bibnamefont
  {B{\"a}uerle}},\ and\ \bibinfo {author} {\bibfnamefont {C.}~\bibnamefont
  {Bechinger}},\ }\bibfield  {title} {\bibinfo {title} {Group formation and
  cohesion of active particles with visual perception{\textendash}dependent
  motility},\ }\href {https://doi.org/10.1126/science.aau5347} {\bibfield
  {journal} {\bibinfo  {journal} {Science}\ }\textbf {\bibinfo {volume}
  {364}},\ \bibinfo {pages} {70} (\bibinfo {year} {2019})}\BibitemShut
  {NoStop}%
\bibitem [{\citenamefont {Curatolo}\ \emph {et~al.}(2020)\citenamefont
  {Curatolo}, \citenamefont {Zhou}, \citenamefont {Zhao}, \citenamefont {Liu},
  \citenamefont {Daerr}, \citenamefont {Tailleur},\ and\ \citenamefont
  {Huang}}]{Curatolo2020Natphys}%
  \BibitemOpen
  \bibfield  {author} {\bibinfo {author} {\bibfnamefont {A.~I.}\ \bibnamefont
  {Curatolo}}, \bibinfo {author} {\bibfnamefont {N.}~\bibnamefont {Zhou}},
  \bibinfo {author} {\bibfnamefont {Y.}~\bibnamefont {Zhao}}, \bibinfo {author}
  {\bibfnamefont {C.}~\bibnamefont {Liu}}, \bibinfo {author} {\bibfnamefont
  {A.}~\bibnamefont {Daerr}}, \bibinfo {author} {\bibfnamefont
  {J.}~\bibnamefont {Tailleur}},\ and\ \bibinfo {author} {\bibfnamefont
  {J.}~\bibnamefont {Huang}},\ }\bibfield  {title} {\bibinfo {title}
  {Cooperative pattern formation in multi-component bacterial systems through
  reciprocal motility regulation},\ }\href
  {https://doi.org/10.1038/s41567-020-0964-z} {\bibfield  {journal} {\bibinfo
  {journal} {Nat. Phys.}\ }\textbf {\bibinfo {volume} {16}},\ \bibinfo {pages}
  {1152} (\bibinfo {year} {2020})}\BibitemShut {NoStop}%
\bibitem [{\citenamefont {Elgeti}\ \emph {et~al.}(2015)\citenamefont {Elgeti},
  \citenamefont {Winkler},\ and\ \citenamefont {Gompper}}]{Elgeti2015review}%
  \BibitemOpen
  \bibfield  {author} {\bibinfo {author} {\bibfnamefont {J.}~\bibnamefont
  {Elgeti}}, \bibinfo {author} {\bibfnamefont {R.~G.}\ \bibnamefont
  {Winkler}},\ and\ \bibinfo {author} {\bibfnamefont {G.}~\bibnamefont
  {Gompper}},\ }\bibfield  {title} {\bibinfo {title} {Physics of
  microswimmers{\textemdash}single particle motion and collective behavior: a
  review},\ }\href {https://doi.org/10.1088/0034-4885/78/5/056601} {\bibfield
  {journal} {\bibinfo  {journal} {Rep. Prog. Phys.}\ }\textbf {\bibinfo
  {volume} {78}},\ \bibinfo {pages} {056601} (\bibinfo {year}
  {2015})}\BibitemShut {NoStop}%
\bibitem [{\citenamefont {Golestanian}(2009)}]{RG:2009}%
  \BibitemOpen
  \bibfield  {author} {\bibinfo {author} {\bibfnamefont {R.}~\bibnamefont
  {Golestanian}},\ }\bibfield  {title} {\bibinfo {title} {Anomalous diffusion
  of symmetric and asymmetric active colloids},\ }\href
  {https://doi.org/10.1103/PhysRevLett.102.188305} {\bibfield  {journal}
  {\bibinfo  {journal} {Phys. Rev. Lett.}\ }\textbf {\bibinfo {volume} {102}},\
  \bibinfo {pages} {188305} (\bibinfo {year} {2009})}\BibitemShut {NoStop}%
\bibitem [{\citenamefont {Bechinger}\ \emph {et~al.}(2016)\citenamefont
  {Bechinger}, \citenamefont {Di~Leonardo}, \citenamefont {L\"owen},
  \citenamefont {Reichhardt}, \citenamefont {Volpe},\ and\ \citenamefont
  {Volpe}}]{BechingerRMP2016}%
  \BibitemOpen
  \bibfield  {author} {\bibinfo {author} {\bibfnamefont {C.}~\bibnamefont
  {Bechinger}}, \bibinfo {author} {\bibfnamefont {R.}~\bibnamefont
  {Di~Leonardo}}, \bibinfo {author} {\bibfnamefont {H.}~\bibnamefont
  {L\"owen}}, \bibinfo {author} {\bibfnamefont {C.}~\bibnamefont {Reichhardt}},
  \bibinfo {author} {\bibfnamefont {G.}~\bibnamefont {Volpe}},\ and\ \bibinfo
  {author} {\bibfnamefont {G.}~\bibnamefont {Volpe}},\ }\bibfield  {title}
  {\bibinfo {title} {Active particles in complex and crowded environments},\
  }\href {https://doi.org/10.1103/RevModPhys.88.045006} {\bibfield  {journal}
  {\bibinfo  {journal} {Rev. Mod. Phys.}\ }\textbf {\bibinfo {volume} {88}},\
  \bibinfo {pages} {045006} (\bibinfo {year} {2016})}\BibitemShut {NoStop}%
\bibitem [{\citenamefont {Berg}\ and\ \citenamefont
  {Brown}(1972)}]{BERG1972Nature}%
  \BibitemOpen
  \bibfield  {author} {\bibinfo {author} {\bibfnamefont {H.~C.}\ \bibnamefont
  {Berg}}\ and\ \bibinfo {author} {\bibfnamefont {D.~A.}\ \bibnamefont
  {Brown}},\ }\bibfield  {title} {\bibinfo {title} {Chemotaxis in escherichia
  coli analysed by three-dimensional tracking},\ }\href
  {https://doi.org/10.1038/239500a0} {\bibfield  {journal} {\bibinfo  {journal}
  {Nature}\ }\textbf {\bibinfo {volume} {239}},\ \bibinfo {pages} {500}
  (\bibinfo {year} {1972})}\BibitemShut {NoStop}%
\bibitem [{\citenamefont {Magariyama}\ \emph {et~al.}(1995)\citenamefont
  {Magariyama}, \citenamefont {Sugiyama}, \citenamefont {Muramoto},
  \citenamefont {Kawagishi}, \citenamefont {Imae},\ and\ \citenamefont
  {Kudo}}]{Magariyama1995BiophysJ}%
  \BibitemOpen
  \bibfield  {author} {\bibinfo {author} {\bibfnamefont {Y.}~\bibnamefont
  {Magariyama}}, \bibinfo {author} {\bibfnamefont {S.}~\bibnamefont
  {Sugiyama}}, \bibinfo {author} {\bibfnamefont {K.}~\bibnamefont {Muramoto}},
  \bibinfo {author} {\bibfnamefont {I.}~\bibnamefont {Kawagishi}}, \bibinfo
  {author} {\bibfnamefont {Y.}~\bibnamefont {Imae}},\ and\ \bibinfo {author}
  {\bibfnamefont {S.}~\bibnamefont {Kudo}},\ }\bibfield  {title} {\bibinfo
  {title} {Simultaneous measurement of bacterial flagellar rotation rate and
  swimming speed},\ }\href
  {https://doi.org/https://doi.org/10.1016/S0006-3495(95)80089-5} {\bibfield
  {journal} {\bibinfo  {journal} {Biophys. J.}\ }\textbf {\bibinfo {volume}
  {69}},\ \bibinfo {pages} {2154} (\bibinfo {year} {1995})}\BibitemShut
  {NoStop}%
\bibitem [{\citenamefont {Paxton}\ \emph {et~al.}(2004)\citenamefont {Paxton},
  \citenamefont {Kistler}, \citenamefont {Olmeda}, \citenamefont {Sen},
  \citenamefont {St.~Angelo}, \citenamefont {Cao}, \citenamefont {Mallouk},
  \citenamefont {Lammert},\ and\ \citenamefont {Crespi}}]{Paxton2004ACS}%
  \BibitemOpen
  \bibfield  {author} {\bibinfo {author} {\bibfnamefont {W.~F.}\ \bibnamefont
  {Paxton}}, \bibinfo {author} {\bibfnamefont {K.~C.}\ \bibnamefont {Kistler}},
  \bibinfo {author} {\bibfnamefont {C.~C.}\ \bibnamefont {Olmeda}}, \bibinfo
  {author} {\bibfnamefont {A.}~\bibnamefont {Sen}}, \bibinfo {author}
  {\bibfnamefont {S.~K.}\ \bibnamefont {St.~Angelo}}, \bibinfo {author}
  {\bibfnamefont {Y.}~\bibnamefont {Cao}}, \bibinfo {author} {\bibfnamefont
  {T.~E.}\ \bibnamefont {Mallouk}}, \bibinfo {author} {\bibfnamefont {P.~E.}\
  \bibnamefont {Lammert}},\ and\ \bibinfo {author} {\bibfnamefont {V.~H.}\
  \bibnamefont {Crespi}},\ }\bibfield  {title} {\bibinfo {title} {{Catalytic
  Nanomotors: Autonomous Movement of Striped Nanorods}},\ }\href
  {https://doi.org/10.1021/ja047697z} {\bibfield  {journal} {\bibinfo
  {journal} {J. Am. Chem. Soc.}\ }\textbf {\bibinfo {volume} {126}},\ \bibinfo
  {pages} {13424} (\bibinfo {year} {2004})}\BibitemShut {NoStop}%
\bibitem [{\citenamefont {Dreyfus}\ \emph {et~al.}(2005)\citenamefont
  {Dreyfus}, \citenamefont {Baudry}, \citenamefont {Roper}, \citenamefont
  {Fermigier}, \citenamefont {Stone},\ and\ \citenamefont
  {Bibette}}]{Dreyfus2005Nature}%
  \BibitemOpen
  \bibfield  {author} {\bibinfo {author} {\bibfnamefont {R.}~\bibnamefont
  {Dreyfus}}, \bibinfo {author} {\bibfnamefont {J.}~\bibnamefont {Baudry}},
  \bibinfo {author} {\bibfnamefont {M.~L.}\ \bibnamefont {Roper}}, \bibinfo
  {author} {\bibfnamefont {M.}~\bibnamefont {Fermigier}}, \bibinfo {author}
  {\bibfnamefont {H.~A.}\ \bibnamefont {Stone}},\ and\ \bibinfo {author}
  {\bibfnamefont {J.}~\bibnamefont {Bibette}},\ }\bibfield  {title} {\bibinfo
  {title} {Microscopic artificial swimmers},\ }\href
  {https://doi.org/10.1038/nature04090} {\bibfield  {journal} {\bibinfo
  {journal} {Nature}\ }\textbf {\bibinfo {volume} {437}},\ \bibinfo {pages}
  {862} (\bibinfo {year} {2005})}\BibitemShut {NoStop}%
\bibitem [{\citenamefont {Howse}\ \emph {et~al.}(2007)\citenamefont {Howse},
  \citenamefont {Jones}, \citenamefont {Ryan}, \citenamefont {Gough},
  \citenamefont {Vafabakhsh},\ and\ \citenamefont
  {Golestanian}}]{Howse2007PRL}%
  \BibitemOpen
  \bibfield  {author} {\bibinfo {author} {\bibfnamefont {J.~R.}\ \bibnamefont
  {Howse}}, \bibinfo {author} {\bibfnamefont {R.~A.~L.}\ \bibnamefont {Jones}},
  \bibinfo {author} {\bibfnamefont {A.~J.}\ \bibnamefont {Ryan}}, \bibinfo
  {author} {\bibfnamefont {T.}~\bibnamefont {Gough}}, \bibinfo {author}
  {\bibfnamefont {R.}~\bibnamefont {Vafabakhsh}},\ and\ \bibinfo {author}
  {\bibfnamefont {R.}~\bibnamefont {Golestanian}},\ }\bibfield  {title}
  {\bibinfo {title} {Self-motile colloidal particles: From directed propulsion
  to random walk},\ }\href {https://doi.org/10.1103/PhysRevLett.99.048102}
  {\bibfield  {journal} {\bibinfo  {journal} {Phys. Rev. Lett.}\ }\textbf
  {\bibinfo {volume} {99}},\ \bibinfo {pages} {048102} (\bibinfo {year}
  {2007})}\BibitemShut {NoStop}%
\bibitem [{\citenamefont {Corkidi}\ \emph {et~al.}(2008)\citenamefont
  {Corkidi}, \citenamefont {Taboada}, \citenamefont {Wood}, \citenamefont
  {Guerrero},\ and\ \citenamefont {Darszon}}]{Corkidi2008Biochem}%
  \BibitemOpen
  \bibfield  {author} {\bibinfo {author} {\bibfnamefont {G.}~\bibnamefont
  {Corkidi}}, \bibinfo {author} {\bibfnamefont {B.}~\bibnamefont {Taboada}},
  \bibinfo {author} {\bibfnamefont {C.}~\bibnamefont {Wood}}, \bibinfo {author}
  {\bibfnamefont {A.}~\bibnamefont {Guerrero}},\ and\ \bibinfo {author}
  {\bibfnamefont {A.}~\bibnamefont {Darszon}},\ }\bibfield  {title} {\bibinfo
  {title} {Tracking sperm in three-dimensions},\ }\href
  {https://doi.org/https://doi.org/10.1016/j.bbrc.2008.05.189} {\bibfield
  {journal} {\bibinfo  {journal} {Biochem. Biophys. Res. Commun.}\ }\textbf
  {\bibinfo {volume} {373}},\ \bibinfo {pages} {125} (\bibinfo {year}
  {2008})}\BibitemShut {NoStop}%
\bibitem [{\citenamefont {Thutupalli}\ \emph {et~al.}(2011)\citenamefont
  {Thutupalli}, \citenamefont {Seemann},\ and\ \citenamefont
  {Herminghaus}}]{Thutupalli2011NJP}%
  \BibitemOpen
  \bibfield  {author} {\bibinfo {author} {\bibfnamefont {S.}~\bibnamefont
  {Thutupalli}}, \bibinfo {author} {\bibfnamefont {R.}~\bibnamefont
  {Seemann}},\ and\ \bibinfo {author} {\bibfnamefont {S.}~\bibnamefont
  {Herminghaus}},\ }\bibfield  {title} {\bibinfo {title} {Swarming behavior of
  simple model squirmers},\ }\href
  {https://doi.org/10.1088/1367-2630/13/7/073021} {\bibfield  {journal}
  {\bibinfo  {journal} {New J. Phys.}\ }\textbf {\bibinfo {volume} {13}},\
  \bibinfo {pages} {073021} (\bibinfo {year} {2011})}\BibitemShut {NoStop}%
\bibitem [{\citenamefont {Grosjean}\ \emph {et~al.}(2016)\citenamefont
  {Grosjean}, \citenamefont {Hubert}, \citenamefont {Lagubeau},\ and\
  \citenamefont {Vandewalle}}]{Grosjean2016PRE}%
  \BibitemOpen
  \bibfield  {author} {\bibinfo {author} {\bibfnamefont {G.}~\bibnamefont
  {Grosjean}}, \bibinfo {author} {\bibfnamefont {M.}~\bibnamefont {Hubert}},
  \bibinfo {author} {\bibfnamefont {G.}~\bibnamefont {Lagubeau}},\ and\
  \bibinfo {author} {\bibfnamefont {N.}~\bibnamefont {Vandewalle}},\ }\bibfield
   {title} {\bibinfo {title} {Realization of the najafi-golestanian
  microswimmer},\ }\href {https://doi.org/10.1103/PhysRevE.94.021101}
  {\bibfield  {journal} {\bibinfo  {journal} {Phys. Rev. E}\ }\textbf {\bibinfo
  {volume} {94}},\ \bibinfo {pages} {021101} (\bibinfo {year}
  {2016})}\BibitemShut {NoStop}%
\bibitem [{\citenamefont {Turner}\ \emph {et~al.}(2016)\citenamefont {Turner},
  \citenamefont {Ping}, \citenamefont {Neubauer},\ and\ \citenamefont
  {Berg}}]{Turner2016BioJ}%
  \BibitemOpen
  \bibfield  {author} {\bibinfo {author} {\bibfnamefont {L.}~\bibnamefont
  {Turner}}, \bibinfo {author} {\bibfnamefont {L.}~\bibnamefont {Ping}},
  \bibinfo {author} {\bibfnamefont {M.}~\bibnamefont {Neubauer}},\ and\
  \bibinfo {author} {\bibfnamefont {H.~C.}\ \bibnamefont {Berg}},\ }\bibfield
  {title} {\bibinfo {title} {Visualizing flagella while tracking bacteria},\
  }\href {https://doi.org/https://doi.org/10.1016/j.bpj.2016.05.053} {\bibfield
   {journal} {\bibinfo  {journal} {Biophys. J.}\ }\textbf {\bibinfo {volume}
  {111}},\ \bibinfo {pages} {630} (\bibinfo {year} {2016})}\BibitemShut
  {NoStop}%
\bibitem [{\citenamefont {Fragkopoulos}\ \emph {et~al.}(2022)\citenamefont
  {Fragkopoulos}, \citenamefont {Vachier}, \citenamefont {Frey}, \citenamefont
  {Le~Menn}, \citenamefont {Mazza}, \citenamefont {Wilczek}, \citenamefont
  {Zwicker},\ and\ \citenamefont {B{\"a}umchen}}]{Fragkopoulos2021}%
  \BibitemOpen
  \bibfield  {author} {\bibinfo {author} {\bibfnamefont {A.~A.}\ \bibnamefont
  {Fragkopoulos}}, \bibinfo {author} {\bibfnamefont {J.}~\bibnamefont
  {Vachier}}, \bibinfo {author} {\bibfnamefont {J.}~\bibnamefont {Frey}},
  \bibinfo {author} {\bibfnamefont {F.-M.}\ \bibnamefont {Le~Menn}}, \bibinfo
  {author} {\bibfnamefont {M.~G.}\ \bibnamefont {Mazza}}, \bibinfo {author}
  {\bibfnamefont {M.}~\bibnamefont {Wilczek}}, \bibinfo {author} {\bibfnamefont
  {D.}~\bibnamefont {Zwicker}},\ and\ \bibinfo {author} {\bibfnamefont
  {O.}~\bibnamefont {B{\"a}umchen}},\ }\bibfield  {title} {\bibinfo {title}
  {Self-generated oxygen gradients control collective aggregation of
  photosynthetic microbes},\ }\href {https://doi.org/10.1098/rsif.2021.0553}
  {\bibfield  {journal} {\bibinfo  {journal} {J. R. Soc. Interface}\ }\textbf
  {\bibinfo {volume} {18}},\ \bibinfo {pages} {20210553} (\bibinfo {year}
  {2022})}\BibitemShut {NoStop}%
\bibitem [{\citenamefont {Golestanian}\ \emph {et~al.}(2005)\citenamefont
  {Golestanian}, \citenamefont {Liverpool},\ and\ \citenamefont
  {Ajdari}}]{RG:2005}%
  \BibitemOpen
  \bibfield  {author} {\bibinfo {author} {\bibfnamefont {R.}~\bibnamefont
  {Golestanian}}, \bibinfo {author} {\bibfnamefont {T.~B.}\ \bibnamefont
  {Liverpool}},\ and\ \bibinfo {author} {\bibfnamefont {A.}~\bibnamefont
  {Ajdari}},\ }\bibfield  {title} {\bibinfo {title} {Propulsion of a molecular
  machine by asymmetric distribution of reaction products},\ }\href
  {https://doi.org/10.1103/PhysRevLett.94.220801} {\bibfield  {journal}
  {\bibinfo  {journal} {Phys. Rev. Lett.}\ }\textbf {\bibinfo {volume} {94}},\
  \bibinfo {pages} {220801} (\bibinfo {year} {2005})}\BibitemShut {NoStop}%
\bibitem [{\citenamefont {Najafi}\ and\ \citenamefont
  {Golestanian}(2004)}]{Najafi:2004}%
  \BibitemOpen
  \bibfield  {author} {\bibinfo {author} {\bibfnamefont {A.}~\bibnamefont
  {Najafi}}\ and\ \bibinfo {author} {\bibfnamefont {R.}~\bibnamefont
  {Golestanian}},\ }\bibfield  {title} {\bibinfo {title} {Simple swimmer at low
  reynolds number: Three linked spheres},\ }\href
  {https://doi.org/10.1103/PhysRevE.69.062901} {\bibfield  {journal} {\bibinfo
  {journal} {Phys. Rev. E}\ }\textbf {\bibinfo {volume} {69}},\ \bibinfo
  {pages} {062901} (\bibinfo {year} {2004})}\BibitemShut {NoStop}%
\bibitem [{\citenamefont {Schienbein}\ and\ \citenamefont
  {Gruler}(1993)}]{Schienbein1993BMB}%
  \BibitemOpen
  \bibfield  {author} {\bibinfo {author} {\bibfnamefont {M.}~\bibnamefont
  {Schienbein}}\ and\ \bibinfo {author} {\bibfnamefont {H.}~\bibnamefont
  {Gruler}},\ }\bibfield  {title} {\bibinfo {title} {{Langevin equation,
  Fokker-Planck equation and cell migration}},\ }\href
  {https://doi.org/https://doi.org/10.1016/S0092-8240(05)80241-1} {\bibfield
  {journal} {\bibinfo  {journal} {Bull. Math. Biol.}\ }\textbf {\bibinfo
  {volume} {55}},\ \bibinfo {pages} {585} (\bibinfo {year} {1993})}\BibitemShut
  {NoStop}%
\bibitem [{\citenamefont {Peruani}\ and\ \citenamefont
  {Morelli}(2007)}]{peruaniPRL2007}%
  \BibitemOpen
  \bibfield  {author} {\bibinfo {author} {\bibfnamefont {F.}~\bibnamefont
  {Peruani}}\ and\ \bibinfo {author} {\bibfnamefont {L.~G.}\ \bibnamefont
  {Morelli}},\ }\bibfield  {title} {\bibinfo {title} {Self-propelled particles
  with fluctuating speed and direction of motion in two dimensions},\ }\href
  {https://doi.org/10.1103/PhysRevLett.99.010602} {\bibfield  {journal}
  {\bibinfo  {journal} {Phys. Rev. Lett.}\ }\textbf {\bibinfo {volume} {99}},\
  \bibinfo {pages} {010602} (\bibinfo {year} {2007})}\BibitemShut {NoStop}%
\bibitem [{\citenamefont {Romanczuk}\ and\ \citenamefont
  {Schimansky-Geier}(2011)}]{Romanczuk2011PRL}%
  \BibitemOpen
  \bibfield  {author} {\bibinfo {author} {\bibfnamefont {P.}~\bibnamefont
  {Romanczuk}}\ and\ \bibinfo {author} {\bibfnamefont {L.}~\bibnamefont
  {Schimansky-Geier}},\ }\bibfield  {title} {\bibinfo {title} {Brownian motion
  with active fluctuations},\ }\href
  {https://doi.org/10.1103/PhysRevLett.106.230601} {\bibfield  {journal}
  {\bibinfo  {journal} {Phys. Rev. Lett.}\ }\textbf {\bibinfo {volume} {106}},\
  \bibinfo {pages} {230601} (\bibinfo {year} {2011})}\BibitemShut {NoStop}%
\bibitem [{\citenamefont {Chaudhuri}(2014)}]{Chaudhuri2014PRE}%
  \BibitemOpen
  \bibfield  {author} {\bibinfo {author} {\bibfnamefont {D.}~\bibnamefont
  {Chaudhuri}},\ }\bibfield  {title} {\bibinfo {title} {Active brownian
  particles: Entropy production and fluctuation response},\ }\href
  {https://doi.org/10.1103/PhysRevE.90.022131} {\bibfield  {journal} {\bibinfo
  {journal} {Phys. Rev. E}\ }\textbf {\bibinfo {volume} {90}},\ \bibinfo
  {pages} {022131} (\bibinfo {year} {2014})}\BibitemShut {NoStop}%
\bibitem [{\citenamefont {Caprini}\ \emph
  {et~al.}(2022{\natexlab{a}})\citenamefont {Caprini}, \citenamefont
  {Sprenger}, \citenamefont {L{\"o}wen},\ and\ \citenamefont
  {Wittmann}}]{Caprini2022JCP}%
  \BibitemOpen
  \bibfield  {author} {\bibinfo {author} {\bibfnamefont {L.}~\bibnamefont
  {Caprini}}, \bibinfo {author} {\bibfnamefont {A.~R.}\ \bibnamefont
  {Sprenger}}, \bibinfo {author} {\bibfnamefont {H.}~\bibnamefont
  {L{\"o}wen}},\ and\ \bibinfo {author} {\bibfnamefont {R.}~\bibnamefont
  {Wittmann}},\ }\bibfield  {title} {\bibinfo {title} {The parental active
  model: A unifying stochastic description of self-propulsion},\ }\href
  {https://doi.org/10.1063/5.0084213} {\bibfield  {journal} {\bibinfo
  {journal} {The Journal of Chemical Physics}\ }\textbf {\bibinfo {volume}
  {156}},\ \bibinfo {pages} {071102} (\bibinfo {year}
  {2022}{\natexlab{a}})}\BibitemShut {NoStop}%
\bibitem [{Note1()}]{Note1}%
  \BibitemOpen
  \bibinfo {note} {See Supplementary Material at [url].}\BibitemShut {Stop}%
\bibitem [{\citenamefont {Martin}\ \emph {et~al.}(2021)\citenamefont {Martin},
  \citenamefont {O'Byrne}, \citenamefont {Cates}, \citenamefont {Fodor},
  \citenamefont {Nardini}, \citenamefont {Tailleur},\ and\ \citenamefont {van
  Wijland}}]{Martin2021PRE}%
  \BibitemOpen
  \bibfield  {author} {\bibinfo {author} {\bibfnamefont {D.}~\bibnamefont
  {Martin}}, \bibinfo {author} {\bibfnamefont {J.}~\bibnamefont {O'Byrne}},
  \bibinfo {author} {\bibfnamefont {M.~E.}\ \bibnamefont {Cates}}, \bibinfo
  {author} {\bibfnamefont {E.}~\bibnamefont {Fodor}}, \bibinfo {author}
  {\bibfnamefont {C.}~\bibnamefont {Nardini}}, \bibinfo {author} {\bibfnamefont
  {J.}~\bibnamefont {Tailleur}},\ and\ \bibinfo {author} {\bibfnamefont
  {F.}~\bibnamefont {van Wijland}},\ }\bibfield  {title} {\bibinfo {title}
  {Statistical mechanics of active ornstein-uhlenbeck particles},\ }\href
  {https://doi.org/10.1103/PhysRevE.103.032607} {\bibfield  {journal} {\bibinfo
   {journal} {Phys. Rev. E}\ }\textbf {\bibinfo {volume} {103}},\ \bibinfo
  {pages} {032607} (\bibinfo {year} {2021})}\BibitemShut {NoStop}%
\bibitem [{\citenamefont {Schnitzer}(1993)}]{SchnitzerPRE1993}%
  \BibitemOpen
  \bibfield  {author} {\bibinfo {author} {\bibfnamefont {M.~J.}\ \bibnamefont
  {Schnitzer}},\ }\bibfield  {title} {\bibinfo {title} {Theory of continuum
  random walks and application to chemotaxis},\ }\href
  {https://doi.org/10.1103/PhysRevE.48.2553} {\bibfield  {journal} {\bibinfo
  {journal} {Phys. Rev. E}\ }\textbf {\bibinfo {volume} {48}},\ \bibinfo
  {pages} {2553} (\bibinfo {year} {1993})}\BibitemShut {NoStop}%
\bibitem [{\citenamefont {Fernandez-Rodriguez}\ \emph
  {et~al.}(2020)\citenamefont {Fernandez-Rodriguez}, \citenamefont {Grillo},
  \citenamefont {Alvarez}, \citenamefont {Rathlef}, \citenamefont {Buttinoni},
  \citenamefont {Volpe},\ and\ \citenamefont {Isa}}]{Fernandez2020NatCom}%
  \BibitemOpen
  \bibfield  {author} {\bibinfo {author} {\bibfnamefont {M.~A.}\ \bibnamefont
  {Fernandez-Rodriguez}}, \bibinfo {author} {\bibfnamefont {F.}~\bibnamefont
  {Grillo}}, \bibinfo {author} {\bibfnamefont {L.}~\bibnamefont {Alvarez}},
  \bibinfo {author} {\bibfnamefont {M.}~\bibnamefont {Rathlef}}, \bibinfo
  {author} {\bibfnamefont {I.}~\bibnamefont {Buttinoni}}, \bibinfo {author}
  {\bibfnamefont {G.}~\bibnamefont {Volpe}},\ and\ \bibinfo {author}
  {\bibfnamefont {L.}~\bibnamefont {Isa}},\ }\bibfield  {title} {\bibinfo
  {title} {Feedback-controlled active brownian colloids with space-dependent
  rotational dynamics},\ }\href {https://doi.org/10.1038/s41467-020-17864-4}
  {\bibfield  {journal} {\bibinfo  {journal} {Nat. Commun.}\ }\textbf {\bibinfo
  {volume} {11}},\ \bibinfo {pages} {4223} (\bibinfo {year}
  {2020})}\BibitemShut {NoStop}%
\bibitem [{\citenamefont {Fischer}\ \emph {et~al.}(2020)\citenamefont
  {Fischer}, \citenamefont {Schmid},\ and\ \citenamefont
  {Speck}}]{Fischer2020PRE}%
  \BibitemOpen
  \bibfield  {author} {\bibinfo {author} {\bibfnamefont {A.}~\bibnamefont
  {Fischer}}, \bibinfo {author} {\bibfnamefont {F.}~\bibnamefont {Schmid}},\
  and\ \bibinfo {author} {\bibfnamefont {T.}~\bibnamefont {Speck}},\ }\bibfield
   {title} {\bibinfo {title} {Quorum-sensing active particles with
  discontinuous motility},\ }\href
  {https://doi.org/10.1103/PhysRevE.101.012601} {\bibfield  {journal} {\bibinfo
   {journal} {Phys. Rev. E}\ }\textbf {\bibinfo {volume} {101}},\ \bibinfo
  {pages} {012601} (\bibinfo {year} {2020})}\BibitemShut {NoStop}%
\bibitem [{\citenamefont {Row}\ and\ \citenamefont {Brady}(2020)}]{Row2020PRE}%
  \BibitemOpen
  \bibfield  {author} {\bibinfo {author} {\bibfnamefont {H.}~\bibnamefont
  {Row}}\ and\ \bibinfo {author} {\bibfnamefont {J.~F.}\ \bibnamefont
  {Brady}},\ }\bibfield  {title} {\bibinfo {title} {Reverse osmotic effect in
  active matter},\ }\href {https://doi.org/10.1103/PhysRevE.101.062604}
  {\bibfield  {journal} {\bibinfo  {journal} {Phys. Rev. E}\ }\textbf {\bibinfo
  {volume} {101}},\ \bibinfo {pages} {062604} (\bibinfo {year}
  {2020})}\BibitemShut {NoStop}%
\bibitem [{\citenamefont {Lee}(2017)}]{Lee2017SoftMat}%
  \BibitemOpen
  \bibfield  {author} {\bibinfo {author} {\bibfnamefont {C.~F.}\ \bibnamefont
  {Lee}},\ }\bibfield  {title} {\bibinfo {title} {Interface stability{,}
  interface fluctuations{,} and the gibbs--thomson relationship in
  motility-induced phase separations},\ }\href
  {https://doi.org/10.1039/C6SM01978A} {\bibfield  {journal} {\bibinfo
  {journal} {Soft Matter}\ }\textbf {\bibinfo {volume} {13}},\ \bibinfo {pages}
  {376} (\bibinfo {year} {2017})}\BibitemShut {NoStop}%
\bibitem [{\citenamefont {Solon}\ \emph {et~al.}(2018)\citenamefont {Solon},
  \citenamefont {Stenhammar}, \citenamefont {Cates}, \citenamefont {Kafri},\
  and\ \citenamefont {Tailleur}}]{Solon2018NJP}%
  \BibitemOpen
  \bibfield  {author} {\bibinfo {author} {\bibfnamefont {A.~P.}\ \bibnamefont
  {Solon}}, \bibinfo {author} {\bibfnamefont {J.}~\bibnamefont {Stenhammar}},
  \bibinfo {author} {\bibfnamefont {M.~E.}\ \bibnamefont {Cates}}, \bibinfo
  {author} {\bibfnamefont {Y.}~\bibnamefont {Kafri}},\ and\ \bibinfo {author}
  {\bibfnamefont {J.}~\bibnamefont {Tailleur}},\ }\bibfield  {title} {\bibinfo
  {title} {Generalized thermodynamics of motility-induced phase separation:
  phase equilibria, laplace pressure, and change of ensembles},\ }\href
  {https://doi.org/10.1088/1367-2630/aaccdd} {\bibfield  {journal} {\bibinfo
  {journal} {New J. Phys.}\ }\textbf {\bibinfo {volume} {20}},\ \bibinfo
  {pages} {075001} (\bibinfo {year} {2018})}\BibitemShut {NoStop}%
\bibitem [{\citenamefont {Omar}\ \emph {et~al.}(2020)\citenamefont {Omar},
  \citenamefont {Wang},\ and\ \citenamefont {Brady}}]{Omar2020PRE}%
  \BibitemOpen
  \bibfield  {author} {\bibinfo {author} {\bibfnamefont {A.~K.}\ \bibnamefont
  {Omar}}, \bibinfo {author} {\bibfnamefont {Z.-G.}\ \bibnamefont {Wang}},\
  and\ \bibinfo {author} {\bibfnamefont {J.~F.}\ \bibnamefont {Brady}},\
  }\bibfield  {title} {\bibinfo {title} {Microscopic origins of the swim
  pressure and the anomalous surface tension of active matter},\ }\href
  {https://doi.org/10.1103/PhysRevE.101.012604} {\bibfield  {journal} {\bibinfo
   {journal} {Phys. Rev. E}\ }\textbf {\bibinfo {volume} {101}},\ \bibinfo
  {pages} {012604} (\bibinfo {year} {2020})}\BibitemShut {NoStop}%
\bibitem [{\citenamefont {Bialk{\'{e}}}\ \emph {et~al.}(2013)\citenamefont
  {Bialk{\'{e}}}, \citenamefont {L{\"o}wen},\ and\ \citenamefont
  {Speck}}]{Bialke2013EPL}%
  \BibitemOpen
  \bibfield  {author} {\bibinfo {author} {\bibfnamefont {J.}~\bibnamefont
  {Bialk{\'{e}}}}, \bibinfo {author} {\bibfnamefont {H.}~\bibnamefont
  {L{\"o}wen}},\ and\ \bibinfo {author} {\bibfnamefont {T.}~\bibnamefont
  {Speck}},\ }\bibfield  {title} {\bibinfo {title} {Microscopic theory for the
  phase separation of self-propelled repulsive disks},\ }\href
  {https://doi.org/10.1209/0295-5075/103/30008} {\bibfield  {journal} {\bibinfo
   {journal} {{EPL}}\ }\textbf {\bibinfo {volume} {103}},\ \bibinfo {pages}
  {30008} (\bibinfo {year} {2013})}\BibitemShut {NoStop}%
\bibitem [{\citenamefont {Solon}\ \emph {et~al.}(2015)\citenamefont {Solon},
  \citenamefont {Cates},\ and\ \citenamefont {Tailleur}}]{Solon2015EPJE}%
  \BibitemOpen
  \bibfield  {author} {\bibinfo {author} {\bibfnamefont {A.~P.}\ \bibnamefont
  {Solon}}, \bibinfo {author} {\bibfnamefont {M.~E.}\ \bibnamefont {Cates}},\
  and\ \bibinfo {author} {\bibfnamefont {J.}~\bibnamefont {Tailleur}},\
  }\bibfield  {title} {\bibinfo {title} {Active brownian particles and
  run-and-tumble particles: A comparative study},\ }\href
  {https://doi.org/10.1140/epjst/e2015-02457-0} {\bibfield  {journal} {\bibinfo
   {journal} {Eur. Phys. J. Spec. Top.}\ }\textbf {\bibinfo {volume} {224}},\
  \bibinfo {pages} {1231} (\bibinfo {year} {2015})}\BibitemShut {NoStop}%
\bibitem [{\citenamefont {Wittkowski}\ \emph {et~al.}(2014)\citenamefont
  {Wittkowski}, \citenamefont {Tiribocchi}, \citenamefont {Stenhammar},
  \citenamefont {Allen}, \citenamefont {Marenduzzo},\ and\ \citenamefont
  {Cates}}]{Wittkowski2014AMB}%
  \BibitemOpen
  \bibfield  {author} {\bibinfo {author} {\bibfnamefont {R.}~\bibnamefont
  {Wittkowski}}, \bibinfo {author} {\bibfnamefont {A.}~\bibnamefont
  {Tiribocchi}}, \bibinfo {author} {\bibfnamefont {J.}~\bibnamefont
  {Stenhammar}}, \bibinfo {author} {\bibfnamefont {R.~J.}\ \bibnamefont
  {Allen}}, \bibinfo {author} {\bibfnamefont {D.}~\bibnamefont {Marenduzzo}},\
  and\ \bibinfo {author} {\bibfnamefont {M.~E.}\ \bibnamefont {Cates}},\
  }\bibfield  {title} {\bibinfo {title} {Scalar $\phi$4 field theory for
  active-particle phase separation},\ }\href
  {https://doi.org/10.1038/ncomms5351} {\bibfield  {journal} {\bibinfo
  {journal} {Nat. Commun.}\ }\textbf {\bibinfo {volume} {5}},\ \bibinfo {pages}
  {4351} (\bibinfo {year} {2014})}\BibitemShut {NoStop}%
\bibitem [{\citenamefont {Gro{\ss}mann}\ \emph {et~al.}(2012)\citenamefont
  {Gro{\ss}mann}, \citenamefont {Schimansky-Geier},\ and\ \citenamefont
  {Romanczuk}}]{Grossmann2012}%
  \BibitemOpen
  \bibfield  {author} {\bibinfo {author} {\bibfnamefont {R.}~\bibnamefont
  {Gro{\ss}mann}}, \bibinfo {author} {\bibfnamefont {L.}~\bibnamefont
  {Schimansky-Geier}},\ and\ \bibinfo {author} {\bibfnamefont {P.}~\bibnamefont
  {Romanczuk}},\ }\bibfield  {title} {\bibinfo {title} {Active brownian
  particles with velocity-alignment and active fluctuations},\ }\href
  {https://doi.org/10.1088/1367-2630/14/7/073033} {\bibfield  {journal}
  {\bibinfo  {journal} {New J. Phys.}\ }\textbf {\bibinfo {volume} {14}},\
  \bibinfo {pages} {073033} (\bibinfo {year} {2012})}\BibitemShut {NoStop}%
\bibitem [{\citenamefont {Caprini}\ \emph
  {et~al.}(2022{\natexlab{b}})\citenamefont {Caprini}, \citenamefont {Marconi},
  \citenamefont {Wittmann},\ and\ \citenamefont
  {L{\"o}wen}}]{caprini2022active}%
  \BibitemOpen
  \bibfield  {author} {\bibinfo {author} {\bibfnamefont {L.}~\bibnamefont
  {Caprini}}, \bibinfo {author} {\bibfnamefont {U.~M.~B.}\ \bibnamefont
  {Marconi}}, \bibinfo {author} {\bibfnamefont {R.}~\bibnamefont {Wittmann}},\
  and\ \bibinfo {author} {\bibfnamefont {H.}~\bibnamefont {L{\"o}wen}},\
  }\bibfield  {title} {\bibinfo {title} {{Active particles driven by competing
  spatially dependent self-propulsion and external force}},\ }\href
  {https://doi.org/10.21468/SciPostPhys.13.3.065} {\bibfield  {journal}
  {\bibinfo  {journal} {SciPost Phys.}\ }\textbf {\bibinfo {volume} {13}},\
  \bibinfo {pages} {065} (\bibinfo {year} {2022}{\natexlab{b}})}\BibitemShut
  {NoStop}%
\end{thebibliography}%

\onecolumngrid

\setcounter{equation}{0}
\setcounter{figure}{0}
\renewcommand{\theequation}{S\arabic{equation}}
\renewcommand{\thefigure}{S\arabic{figure}}

\section*{Supplemental Material}

\colrev{
\section{Active noise is multiplicative}

We consider for symmetry reasons a model of active noise for which fluctuations along the norm $v$ and orientation $\n$ of the self-propulsion velocity $\bm v$ are decoupled. 
Such a model may be formulated in terms of separate Langevin equations for $v$ and $\n$, as is done in the main text (see Eqs.~(1b,c)).
Alternatively, an equivalent dynamics can be expressed for the full self-propulsion velocity $\bm v$.
Discarding the positional dynamics, we find after some algebra that the Fokker-Planck equation~(2) of the main text can be re-expressed in Cartesian coordinates as
\begin{equation} \label{eq_FP_mult}
    \partial_t \calP(\bm v,t) = \partial_{v_i}\left[ 
     W_{\rm eff}'(v) \hat{n}_i\calP(\bm v,t) 
    + \Sigma_{il}(\bm v)\Sigma_{kl}(\bm v)\partial_{v_k}\calP(\bm v,t) 
    \right],
\end{equation}
where summation over repeated indices is implied and $\bm \Sigma(\bm v) = \sqrt{D_v}\n\n + v\sqrt{D_r}\bm P^{\perp}(\n)$.
Due to the decoupling between speed and orientation fluctuations, respectively set by the coefficients $D_v$ and $D_r$, the diffusion term in Eq.~\eqref{eq_FP_mult} is multiplicative.
Consequently, the Langevin equation associated with~\eqref{eq_FP_mult} will generally include a noise drift term which will be dependent on the chosen interpretation of the noise.
Using the Stratonovich interpretation, we indeed find that $\bm v$ satisfies
\begin{align}
    \dot{v}_i & = - W'_{\rm eff}(v) \hat{n}_i 
    + \Sigma_{il}(\bm v)\partial_{v_k}\Sigma_{kl}(\bm v)
    + \sqrt{2} \, \Sigma_{il}(\bm v) \,\xi_l(t),\nonumber \\
    \label{eq_Lang_mult}
     & = -\left( W'(v) + (d-1)\sqrt{D_v D_r} \right) \hat{n}_i  + \sqrt{2} \, \Sigma_{il}(\bm v) \, \xi_l(t),
\end{align}
where the Gaussian noise $\bm \xi(t)$ satisfies $\langle \bm \xi(t) \rangle = \bm 0$ and $\langle \xi_i(t) \xi_j(t')\rangle = \delta_{ij}\delta(t - t')$.

}

\section{Langevin simulations}

Brownian dynamics simulations were carried out in three dimensions considering the It\^o SDE equivalent to Eqs.~(1) of the main text, 
namely (including a uniform positional dynamics):
\begin{subequations}
\label{eq_Ito_SDE}
\begin{align}
\dot{\bm r} & = \bm v + \sqrt{2 D_t} \,\bm \xi_t(t) , \\
\dot{\bm v} & = - \left[ \partial_v W(\bm r,v) + (d-1) D_r(\bm r) v \right] \hat{\bm n} 
+\sqrt{2} \left( \sqrt{D_v(\bm r)}\hat{\bm n}\hat{\bm n} + v\sqrt{D_r(\bm r)}\bm P^\perp(\hat{\bm n})  \right) \cdot \bm \xi_v(t) ,
\end{align}
\end{subequations}
where $\bm \xi_t$ and $\bm \xi_v$ and independent Gaussian white noises and for simplicity $D_t$ is taken constant here.
Eqs.~\eqref{eq_Ito_SDE} were integrated using an explicit Euler-Maruyama scheme and
all simulations were done with a time step increment $\rmd t = 10^{-3}$.

To study the effect of spatially varying activity on the stationary state of the system we considered one-dimensional profiles along the $\hat{\bm x}$ direction
splitting the space in two regions with different values of $v_0$ and/or $D_v$, namely
\begin{align*}
v_0(x) & = v_{1} + \frac{v_{2} - v_{1}}{2}\left[ \tanh\left( \frac{x - \tfrac{L}{4}}{w} \right) - \tanh\left( \frac{x - \tfrac{3L}{4}}{w} \right) \right] , \\
D_v(x) & = D_{v,1} + \frac{D_{v,2} -  D_{v,1}}{2}\left[ \tanh\left( \frac{x - \tfrac{L}{4}}{w} \right) - \tanh\left( \frac{x - \tfrac{3L}{4}}{w} \right) \right] ,
\end{align*}
with $L = 100$ the total linear system size and $w$ setting the width of the interfaces separating the two regions. 
Table~\ref{tab_numerics} summarizes the different values of the parameters we used in simulations.
All numerical simulation data shown in Figs.~1 and 2 of the main text are averaged over time, independent trajectories and the remaining two space directions. 

\begin{table}[t!]
\caption{\label{tab_numerics} Summary of the parameter values corresponding to the numerical results presented in the main text.} 
\begin{ruledtabular}
\begin{tabular}{c c c c c c}
   & $v_1$ & $v_2$ & $D_{v,1}$ & $D_{v,2}$ & $w$\\
  Fig. 1(a) & 1 & 20 & 0.1 & 0.1 & 10\\
  Fig. 1(b) & 0.1 & 0.1 & 1 & 20 & 10\\
  Figs. 1(c-f) & 1 & 2 & $\in [1;5]$ (see caption) & 0.1 & 5\\
  Fig. 2 & 1 & 2 & 0.1 & 0.1 & 5\\
\end{tabular}
\end{ruledtabular}
\end{table}

\section{Boundary terms in the moments equations}

In this section we provide details about the derivation of the boundary terms in the equations for the speed and orientation moments.
For convenience, we recall here the Fokker-Planck equation ruling the dynamics of $\calP(\bm r,v,\n,t)$ (Eq.~(2) of the main text):
\begin{equation}
\partial_t {\cal P}(\bm r,v,\hat{\bm n},t) = - \nabla \cdot {\bm J_{\bm r}} + \frac{1}{v^{d-1}} \partial_v\left[v^{d-1} \left( \partial_v W_{\rm eff}(\bm r,v) + D_v(\bm r) \partial_v \right){\cal P}(\bm r,v,\hat{\bm n},t) \right] 
+ D_r(\bm r)\nabla_{\hat{\bm n}}^2 {\cal P}(\bm r,v,\hat{\bm n},t) \,,
\label{eq_app_Kramers} 
\end{equation}
where $W_{\rm eff}(\bm r,v)= W(\bm r,v) + (d-1)D_v(\bm r)\ln v$, and $\bm J_{\bm r} = (\bm v - D_t\nabla){\cal P}$.

For the purpose of this discussion, ${\cal U}(\hat{\bm n})$ denotes an arbitrary function of the director $\hat{\bm n}$.
Multiplying Eq.~\eqref{eq_app_Kramers} by $v^k {\cal U}(\hat{\bm n})$ and integrating over $\bm v$, 
only the speed current $-v^{1-d}\partial_v [v^{d-1}(\partial_v W_{\rm eff} + D_v \partial_v){\cal P}(\bm r, v, \hat{\bm n},t)]$ leads to non-vanishing boundary terms.
Hence, we are interested in evaluating the integral 
\begin{align} 
& \int {\rm d}\hat{\bm n} {\cal U}(\hat{\bm n}) \int {\rm d}v \,v^{k} \partial_v \left[v^{d-1}(\partial_v W_{\rm eff} + D_v \partial_v){\cal P}(\bm r, v, \hat{\bm n},t)\right] 
 = k \rho \langle {\cal U}(\hat{\bm n})[ (k-1)D_v v^{k-2} - v^{k-1} \partial_v W ] \rangle \nonumber \\
 \label{eq_boundary_def}
&\qquad\qquad\qquad\qquad + \lim_{v\to 0} \int{\rm d}\hat{\bm n} {\cal U}(\hat{\bm n}) v^{k+d-1} \left[ k D_v v^{-1} - \partial_v W_{\rm eff} - D_v \partial_v \right] {\cal P}(\bm r, v, \hat{\bm n},t)  , 
\end{align}
where the rhs was obtained after successive integration by parts over $v$, 
and the average $\rho(\bm r,t)\langle \cdot \rangle \equiv \int {\rm d}^d{\bm v} \, (\cdot) {\cal P}({\bm r},v,\n,t)$ is the same as defined in the main text.
As ${\cal P}(\bm r, v, \hat{\bm n},t)$ may diverge in the limit $v \to 0$ (see Eq. (3) of the main text), 
the boundary term on the second line of Eq.~\eqref{eq_boundary_def} is generally nonzero.
Its expression however depends on the full distribution ${\cal P}(\bm r, v, \hat{\bm n},t)$, so that in most cases it is not possible to calculate it exactly.
We now show how it can be derived from a perturbative expansion in spatial gradients.

We recall that, in principle, all parameters of the problem can be space dependent, in which case the distribution in phase space 
${\cal P}(\bm r, v, \hat{\bm n},t)$ cannot be factorized.
In large systems and over long times however, the particle density $\rho$ is the only slow mode of the dynamics, so that we can write
${\cal P}(\bm r, v, \hat{\bm n},t) = \rho(\bm r,t)Q(\bm r,v,\hat{\bm n}|\rho)$, where the distribution $Q$ fixes the speed and orientation moments and solves at fixed $\rho$ the equation 
(from~\eqref{eq_app_Kramers})
\begin{equation} \label{eq_Q}
Q  \nabla\cdot(\rho \langle v \hat{\bm n} \rangle ) - \nabla\cdot(\rho v \hat{\bm n} Q) + D_t\nabla\rho \cdot \nabla Q + \nabla \cdot (D_t \rho \nabla Q) + \rho D_r \nabla_{\hat{\bm n} }^2 Q
- \rho v^{1-d}\partial_v \left[v^{d-1}(\partial_v W_{\rm eff} + D_v\partial_v)Q \right] = 0 .
\end{equation}
In the main text we have shown that nonzero gradients lead to local orientational order,  
we therefore propose the following ansatz
\begin{equation} \label{eq_ansatz_PS}
{Q}(\bm r, v, \hat{\bm n}|\rho) \equiv \frac{1}{{\cal S}_d} \left[ F_0(\bm r, v|\rho) + (\hat{\bm n}\cdot \nabla) F_1(\bm r, v|\rho)  + {\cal O}(\nabla^2) \right],
\end{equation}   
where ${\cal S}_d$ denotes the surface of the unit ($d-1$)-sphere embedded in $d$ dimensions, and we have stopped the expansion at first order in spatial gradients.
Replacing this expression into Eq.~\eqref{eq_boundary_def}, it leads for isotropic (${\cal U}(\hat{\bm n}) = 1$) and polarization (${\cal U}(\hat{\bm n}) = \hat{\bm n}$)
moments, respectively to
\begin{align*}
& \lim_{v\to 0} \int{\rm d}\hat{\bm n}\, v^{k+d-1} \left[ k D_v v^{-1} - \partial_v W_{\rm eff} - D_v \partial_v \right] {\cal P}(\bm r, v, \hat{\bm n},t)  
= \lim_{v\to 0} v^{k+d-1} \left[ k D_v v^{-1} - \partial_v W_{\rm eff} - D_v \partial_v \right] \rho(\bm r,t) F_0(\bm r, v|\rho), \\
& \lim_{v\to 0} \int{\rm d}\hat{\bm n}\,  \hat{\bm n} v^{k+d-1} \left[ k D_v v^{-1} - \partial_v W_{\rm eff} - D_v \partial_v \right] {\cal P}(\bm r, v, \hat{\bm n},t)  
= \lim_{v\to 0} v^{k+d-1} \left[ k D_v v^{-1} - \partial_v W_{\rm eff} - D_v \partial_v \right] \frac{\rho(\bm r,t) \nabla F_1(\bm r, v|\rho)}{d}.
\end{align*}
so that calculating the boundary terms reduces to finding the expressions of $F_0$ and $F_1$ in the limit $v \to 0$.
Replacing the ansatz~\eqref{eq_ansatz_PS} into Eq.~\eqref{eq_Q}, keeping only leading order terms in gradients and equating terms of same order
we find that $F_0$ and $F_1$ must satisfy
\begin{subequations}
\begin{align} 
& (\partial_v W_{\rm eff} + D_v \partial_v) F_0(\bm r, v|\rho) = 0 , \\
& v^{1-d}\partial_v \left[v^{d-1}(\partial_v W_{\rm eff} + D_v \partial_v) (\hat{\bm n}\cdot \nabla) F_1(\bm r, v|\rho) \right] - (d-1)D_r (\hat{\bm n}\cdot \nabla) F_1(\bm r, v|\rho) = v (\hat{\bm n}\cdot \nabla) F_0(\bm r, v|\rho) .
\end{align}
\end{subequations}
The first equation corresponds to vanishing current `along' $v$, it therefore gives 
\begin{equation} \label{eq_F0}
F_0(\bm r, v|\rho) =  \frac{e^{-W_{\rm eff}(\bm r,v)/D_v(\bm r)}}{ Z(\bm r)} = \frac{e^{-W(\bm r,v)/D_v(\bm r)}}{v^{d-1} Z(\bm r)} =  {\cal P}_{\rm H}(\bm r,v),
\end{equation} 
which is the speed distribution solving the spatially homogeneous problem (Eq.~(3) of the main text).
To address the second equation, 
we consider the rescaled distribution $Y(\bm r,v,\hat{\bm n}|\rho) \equiv Z(\bm r) \exp[W_{\rm eff}(\bm r,v)/D_v(\bm r)] (\hat{\bm n}\cdot \nabla) F_1(\bm r, v|\rho)$,
which solves
\begin{equation} \label{eq_reduced_Y} 
\left[ D_v \partial^2_{vv} - (\partial_v W)\partial_v - (d-1)D_r  \right] Y(\bm r, v,\hat{\bm n}|\rho) 
= v(\hat{\bm n}\cdot \nabla)  \ln\left(\frac{\rho}{Z} e^{-W/D_v} \right) .
\end{equation}
Eq.~\eqref{eq_reduced_Y} in general does not admit analytical solutions, even for the simple quadratic potential $W$ used in the main text.
We nevertheless note that in the small active noise regime $(|W| \gg D_v)$ the boundary terms vanish exponentially fast with $W/D_v$ (see e.g.\ Eq.~\eqref{eq_F0}), 
so that their effect should be appreciable only in the opposite limit of large $D_v$. 
Considering $D_v \gg W$, we drop the second term on the lhs of Eq.~\eqref{eq_reduced_Y} and set $e^{-W/D_v} = {\cal O}(1)$,
the corresponding solution reads
\begin{equation}
Y(\bm r, v,\hat{\bm n}|\rho) 
\simeq -\frac{v}{(d-1)D_r}(\hat{\bm n}\cdot \nabla)  \ln\left(\frac{\rho}{Z}e^{-W/D_v} \right) ,
\end{equation}
and where the boundary conditions were chosen so that $Y=0$ in the homogeneous case.

Putting all these results together, defining $W(\bm r,0)\equiv W_0$ and $\gamma(\bm r) \equiv \exp(-W_0/D_v)/Z$, we thus finally obtain
\begin{subequations}
\begin{align}
& \lim_{v\to 0} \int{\rm d}\hat{\bm n}\, v^{k+d-1} \left[ kD_v v^{-1} - \partial_v W_{\rm eff} - D_v \partial_v \right] {\cal P}(\bm r, v, \hat{\bm n},t)  
= D_v \rho(\bm r,t) \gamma(\bm r) \delta_{k,1}, \\
& \lim_{v\to 0} \int{\rm d}\hat{\bm n}\,  \hat{\bm n} v^{k+d-1} \left[ kD_v v^{-1} - \partial_v W_{\rm eff} - D_v \partial_v \right] {\cal P}(\bm r, v, \hat{\bm n},t)  
= \frac{D_v}{d(d-1)D_r}\nabla \left[ \rho(\bm r,t) \gamma(\bm r) \right] \delta_{k,0} .
\end{align}
\end{subequations}

\section{The local nematic order}

Following the notations of the main text, 
the equation describing the dynamics of the second orientational moments, i.e.\ those proportional to the nematic order ${\bm q} = \hat{\bm n}\hat{\bm n} - {\bm I}/d$,
is for $k \ge 0$
\begin{equation} \label{eq_qk}
\partial_t \left(\rho \langle v^k \bm q \rangle\right) = - \nabla \cdot \left[ \rho \langle v^{k+1} \bm T \rangle - D_t \nabla \left( \rho \langle v^k \bm q \rangle \right) \right] 
- \frac{2}{d+2} \left[ \nabla \left(\rho \langle v^{k+1} \hat{\bm n} \rangle \right)\right]_{ST} - 2d D_r \rho \langle v^k \bm q \rangle + \rho \langle {\cal G}_k \bm q \rangle + {\cal O}(\nabla^2),
\end{equation}
where $\left([\bm A]_{ST}\right)_{ij} \equiv \tfrac{1}{2} \left( A_{ij} + A_{ji} \right) - \tfrac{1}{d}\delta_{ij} {\rm Tr}(\bm A)$ 
stands for the symmetrized and traceless version of the second rank tensor $\bm A$, 
while $T_{ijl} \equiv \hat{n}_i\hat{n}_j\hat{n}_l - \tfrac{1}{d+2} \left( \delta_{jl}\hat{n}_i + \delta_{il}\hat{n}_j + \delta_{ij}\hat{n}_l \right)$ 
is the symmetrized third order orientational moment. 
The boundary terms normally present on the rhs of Eq.~\eqref{eq_qk} (see the previous section) are accounted for by the ${\cal O}(\nabla^2)$ term.
For the following discussion we assume them small, so that their contribution can be neglected.

Setting $\partial_t \rho \langle v^k \bm q \rangle = 0$ and discarding the third order and positional diffusion terms in Eq.~\eqref{eq_qk} 
we find that the local nematic order corresponding to $k = 0$ is given at leading order by
\begin{equation} \label{eq_q}
\rho \langle \bm q \rangle = -\frac{\tau_r(d-1)}{d(d+2)}\left[ \nabla \left(\rho \langle v \hat{\bm n} \rangle \right)\right]_{ST} .
\end{equation} 
As we show in the main text that $\langle v^k \hat{\bm n} \rangle = {\cal O}(\nabla)$, we indeed find that $\langle \bm q \rangle = {\cal O}(\nabla^2)$.
It is moreover straightforward to show that the moments $\rho \langle v^k \bm q \rangle$ with $k > 0$ are of same order in gradients as $\rho \langle \bm q \rangle$, 
which justifies {\it a posteriori} the fact that we discarded the nematic order contribution when enslaving the polarization moments.
Using the steady state condition $\bm J = \rho \langle v \hat{\bm n} \rangle -  D_t\nabla \rho = \bm 0$, we furthermore find that 
\begin{equation} \label{eq_q_ss}
\rho \langle \bm q \rangle_s = -\frac{\tau_r(d-1)}{d(d+2)}\left[ \nabla \left( D_t \nabla \rho_s \right) \right]_{ST} .
\end{equation} 
Therefore, contrary to the steady state polarization (see Eq.~\eqref{eq_n_general} and Figs.~2(c,d) of the main text)
the direction of the emergent nematic order in stationary state is fully determined by that of density gradients.

\colrev{

\section{Speed-orientation correlations and emergent polar order}

In the main text the expressions of the speed-orientation correlation and polarization at interfaces are given in the low noise regime.
In this section, we provide their full expressions for completeness.
From the Fokker-Planck equation~\eqref{eq_app_Kramers} it is straightforward to show that the density current $\bm J$ is exactly given by
\begin{equation}
\bm J = \rho \langle v \n\rangle - D_t \nabla\rho .
\end{equation}
Using the expression of the mean polarity 
\begin{equation}
\rho \langle \hat{\bm n} \rangle = -\frac{\tau_r}{d}\nabla (\rho  \langle v \rangle) + \frac{D_v\tau_r^2}{d}\nabla \left(\rho \gamma \right),
\end{equation}
the speed orientation correlation function $\bm C(v,\n) =  \langle v \n\rangle - \langle v\rangle \langle \n \rangle$
thus reads
\begin{equation} \label{eq_C_J_SM}
\rho \bm C(v,\n) = \bm J + D_t \nabla\rho + \frac{\tau_r}{d} \langle v\rangle \left[ \nabla (\rho \langle v\rangle) - \alpha \frac{D_v}{\mu}\nabla (\rho \gamma) \right]
\end{equation} 

In steady states characterized by $\bm J = \bm 0$ Eq.~\eqref{eq_C_J_SM} thus implies that
\begin{align}
 \bm C(v,\n) & = \left[D_t + \frac{\tau_r}{d} \langle v\rangle\left( \langle v\rangle - \alpha \frac{D_v}{\mu}\gamma \right) \right]\nabla \ln\rho_s 
 + \frac{\tau_r}{d} \langle v\rangle \left( \nabla \langle v\rangle - \alpha \frac{D_v}{\mu}\nabla \gamma \right), \nonumber \\
  \label{eq_C_SM}
 & = \frac{\tau_r}{d} \left[ \langle v\rangle \left( \nabla \langle v\rangle - \alpha \frac{D_v}{\mu}\nabla \gamma \right) - \frac{d D_t \tau_r^{-1} +  \langle v \rangle^2 - \alpha  \langle v \rangle\gamma D_v/\mu}{ d D_t (1+\alpha)\tau_r^{-1} +  \langle v^2 \rangle + \alpha v_0  \langle v\rangle} 
 \left( \nabla\langle v^2\rangle + \alpha v_0 \nabla\langle v\rangle \right)\right],
\end{align}
where we used the expression of the steady state density profile $\rho_s$ given in Eq. (6) of the main text.
Equation~\eqref{eq_C_SM} shows that spatially varying $\langle v\rangle$ and $\langle v^2 \rangle$ both lead to emergent correlations between speed and polarization in steady state. 
It is moreover straightforward to show that in the constant speed ABPs limit of large $\mu$ the correlation function decays as $|\bm C(v,\n)| \sim \mu^{-1}$.

We now turn to the derivation of the steady state polarization, which is given by
\begin{equation}
\langle \n \rangle_s = -\frac{\tau_r}{d} \left[ \left(\langle v\rangle - \alpha \frac{D_v}{\mu}\gamma \right)\nabla \ln\rho_s + \nabla\langle v\rangle - \alpha \frac{D_v}{\mu}\nabla \gamma  \right],
\end{equation}  
such that after replacing $\rho_s$ by its expression we obtain
\begin{equation} \label{eq_n_general}
\langle \n \rangle_s = \frac{\tau_r}{d} \left \{ \frac{[\langle v\rangle(1-\alpha) + \alpha v_0] \nabla \langle v^2\rangle - \left[ d D_t (1+\alpha)\tau_r^{-1} +  \langle v^2 \rangle 
+ \alpha v_0(\langle v \rangle - v_0)\right]\nabla \langle v \rangle}{d D_t (1+\alpha)\tau_r^{-1} +  \langle v^2 \rangle + \alpha v_0  \langle v\rangle} + \alpha \frac{D_v}{\mu}\nabla \gamma \right\}.
\end{equation} 
Here too it is clear that while gradients of $\langle v \rangle$ will lead to a steady state polarization oriented towards slow (generally dense) regions, those of $\langle v^2 \rangle$ will lead to the opposite trend as exemplified in the main text.
}

\end{document}